\documentclass[useAMS,usenatbib]{mn2e}

 \usepackage{txfonts} 
 \usepackage{natbib} 
 \usepackage{graphicx} 

\title[Exploring the gas phase of distant galaxies]{A decrease of the gas exchanges between galaxies and the IGM, from 12 to 6 billion years ago}

\author[M. Rodrigues et al.] {M. Rodrigues$^{1,2,3}$\thanks{E-mail:
marodrig@eso.org} M. Puech$^{1}$, F. Hammer$^{1}$, B. Rothberg$^{4,5}$, H. Flores$^{1}$\\
$^{1}$European Southern Observatory, Alonso de Cordova 3107 - Casilla 19001 - Vitacura -Santiago, Chile\\
$^{2}$GEPI , Observatoire de Paris, CNRS, University Paris Diderot ; 5 Place Jules Janssen,  92195 Meudon, France\\
$^{3}$CENTRA, Instituto Superior Tecnico, Av. Rovisco Pais 1049-001 Lisboa , Portugal \\
$^{4}$George Mason University, Department of Physics \& Astronomy, MS 3F3, 4400 University Drive, Fairfax, VA 22030, USA\\
$^{5}$Space Telescope Science Institute, 3700 San Martin Drive, Baltimore, MD 21218, USA\\
}

\begin{document}

\date{In press 6 February 2012}

\pagerange{\pageref{firstpage}--\pageref{lastpage}} \pubyear{2011}

\maketitle

\label{firstpage}

\begin{abstract}

Using a representative sample of 65 intermediate mass galaxies at $z\sim0.6$, 
we have investigated the interplay between the main ingredients of  
chemical evolution: metal abundance, gas mass, stellar mass and SFR. 
All quantities have been estimated using deep spectroscopy and photometry 
from UV to IR and assuming an inversion of the Schmitt-Kennicutt law 
for the gas fraction. Six billion years ago, galaxies had a mean gas 
fraction of $32\% \pm 3$, i.e. twice that of their local counterparts. 
Using higher redshift samples from the literature, we explore the 
gas-phases and estimate the evolution of the mean gas fraction
of distant galaxies over the last 11 Gy. The gas fraction increases 
linearly at the rate of 4\% per Gyr from $z\sim0$ to $z\sim2.2$. 
We also demonstrate for a statistically representative sample
that $<$ 4\% of the $z\sim0.6$ galaxies 
are undergoing outflow events, in sharp contrast with $z\sim2.2$ galaxies. 
The observed co-evolution of metals and gas over the past 6 Gyr 
favours a scenario in which the population of intermediate mass 
galaxies evolved as closed-systems, converting 
their own gas reservoirs into stars.
\end{abstract}


\section{Introduction}

In the framework of the $\Lambda$CDM model, galaxies are formed by merging together over time. Although cosmological simulations predict the assembly of dark matter haloes well, little is known about how the baryonic matter is assembled into galaxies inside the dark matter halos . One major challenge in astrophysics is to understand the formation of local disks which make up the bulk of the stellar mass at z=0. Since the first studies on the formation of discs  \citep{1962ApJ...136..748E}, two scenarios have emerged to explain their formation. The first is a secular scenario driven by quasi-adiabatic processes, in which disks have gradually formed by gas accretion. The second is a hierarchical scenario in which the growth of disc galaxies is driven by violent processes such as gas-rich major mergers. The first step in selecting between these two scenarios is to put observational constraints on the relative contributions of mergers, infall of gas, feedback and star formation processes as a function of time to the stellar mass assembly. 

Studies on the evolution of the star formation rate \citep{1999ApJ...517..148F, 2001ApJ...556..562C} and of the stellar mass densities \citep[e.g.]{2001MNRAS.326..255C, 2004ApJ...608..742D} have revealed that in present-day galaxies, 30 to 50\% of the baryonic mass is locked up in stars which formed since $z<1$. This evolution is mostly associated with the evolution of intermediate mass galaxies, with stellar mass between $1.45 -15\times10^{10}\,M_\odot$ \citep{2005A&A...430..115H,2005ApJ...625...23B}. As disk galaxies represent 70\% of intermediate mass galaxies in the local Universe, intermediate mass galaxies at $z\sim0.6$ appear to be the likeliest progenitors of the present day spiral galaxies. The ESO large program IMAGES "Intermediate MAss Galaxy Evolution Sequences" investigated the evolution of the main global properties of the past 6 Gyr using a representative sample of local spiral progenitors at $z\sim0.6$. The study of the fundamental quantities of the IMAGES sample provided strong evidence for the evolution of intermediate-mass galaxies at z=0.7 in terms of morphology \citep{2008A&A...484..159N,2010A&A...509A..78D}, kinematics \citep{2008A&A...477..789Y,2008A&A...484..173P}, star formation and metal content \citep[hereafter R08]{2008A&A...492..371R}. It revealed that nearly half of them are systems not yet dynamically relaxed. In \citet{2009A&A...507.1313H}, we illustrated how these observations are compatible with a galaxy formation scenario in which the present-day galaxies are relics of  major mergers. It does not imply a merger origin for the Hubble sequence but proves that it is a plausible channel for the formation of large local disks. The confirmation of a scenario of galaxy formation will need a complete characterisation of the properties of galaxies at different redshifts and the full comprehension of the physics of star formation. 

In this context, the interstellar medium is the ideal laboratory to test the different galaxy formation scenarios, since it allows us to investigate the chemical cycle of galaxies which behaves like a clock for star-formation. The chemical evolution of galaxies can be defined by four observable parameters: (1) the gas fraction $f_{gas}$; (2) the stellar mass $M_*$; (3) the metallicity of the gas $Z_g$ and  (4) the star formation rate $SFR$. In the case of a galaxy in total isolation, the relation between gas, stellar mass and metallicity are known and the evolution of each quantity can be predicted by the analytical models of chemical enrichment \citep{1990MNRAS.246..678E}. However, in practice galaxies are not isolated systems since they can exchange gas and metals with the environment, e.g: outflow of gas ejected by violent stellar winds, infall of pristine gas from cosmic filaments, accretion and ejection of gas and metals during a merger event. The contribution of these processes on galaxy evolution can be constrained by studying the evolution with time of the four parameters of chemical evolution. In the last five years, a large amount of galaxy surveys have collected gas metallicities and gas fractions in galaxy from low Universe to very high redshift, opening the study of the metal and gas content evolution over a large lookback time interval, e.g \citet{2010arXiv1006.4877J, 2006ApJ...647..128E,2009MNRAS.398.1915M}. These studies have put in evidence the importance of outflow and infalls at $z>2$ on the evolution of galaxies \citep{2008ApJ...674..151E,2009MNRAS.398.1915M}.  Nevertheless, these environmental processes add at least four supplementary parameters in chemical evolution models: the mass of gas and of metals from infall $M_{g,inf}$ and $Z_{g,inf}$ and the mass of gas and metals expelled by outflows $M_{g,out}$ and $Z_{g,out}$. The addition of these parameters makes the problem degenerate, because the physical description of outflows and infalls and their dependences with $M_*$, $SFR$ are unknown. In order to compare the analytical models of chemical evolution with observations, strong assumptions need to be made about the characteristics of environmental processes, in particular on the metallicity of the in or out- falling gas. Recenly, \citet{2011arXiv1110.0837R, 2011ApJ...728...55R, 2010Natur.467..811C} (among other) have directly detected infall and outflows in distant intermediate mass galaxies. These observations provide the first direct constraints on the properties of the outflows and infalls, such as the metallicity, spatial extent and morphology of the out-in-falling gas. Estimates of these parameters are crucial for understanding the impact of outflows/nfall in driving galaxy evolution.

In this paper, we extend the study of the evolution of the fundamental
quantities of disc progenitors in the framework of the IMAGES
survey, to the observables of the interstellar medium (ISM). In R08,
we have gathered a complete sample of 88 star-forming galaxies
at $z\sim0.6$ and estimated their gas metallicity. We used a closed-box chemical model to infer the evolution of the gas fraction over the past 6 Gyr. In the present paper, we use both the literature and the Schmidt-Kennicutt law to estimate the evolution of the gas fraction over the past 11 Gyr, taking care to homogenise the data. We then investigate if the co-evolution of both the gas fraction and gas metallicity in the population of intermediate-mass galaxies as a function of time is consistent with a simple closed-box chemical model.  This paper is organised as follows. Section 2 explains the sample selection and the estimation of the main observables of chemical evolution - stellar mass, metallicity, SFR and gas fraction. In Section 3, we characterise the gas phase of a representative sample of intermediate mass galaxies at z $\sim$0.7 in terms of the ionisation properties of HII regions, gas fraction and the presence of outflows. In Section 4 we extend the constraints on the evolution of the gas fraction to larger lookback times in the population of intermediate mass galaxies using estimates from the literature. In section 5, we combine the gas fraction evolution with the evolution of metals derived in R08 and compare that to the prediction of the closed-box model.Throughout this paper we adopt a $\Lambda$-CDM cosmological model of $\mathrm{H_0}$=70$\mathrm{km~s^{-1}Mpc^{-1}}$, $\Omega_\mathrm{M}$=0.3 and $\Omega_\mathrm{\Lambda}$=0.7. All magnitudes used in this paper are in the AB system, unless explicitly noted otherwise. The adopted solar abundance is $12+log(O/H)$=8.66 from \citet{2004A&A...417..751A} and a nucleosynthetic yield y=0.0126 \citep{2004A&A...417..751A}.

\section{Sample and data analysis}

\subsection{Sample selection}
\begin{figure}
\resizebox{\hsize}{!}{\includegraphics{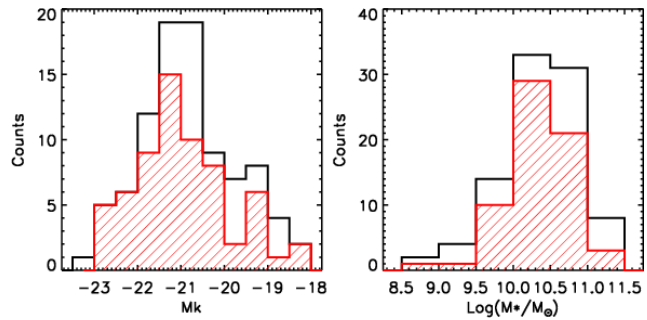}}
\caption{ Distribution of absolute magnitude in K-band (upper-left) and stellar mass (down-right) for all star-forming galaxies ($EW([OII])>15\AA$) in the representative sample of intermediate mass galaxies IMAGES-FORS2 (black histogram) and for our sample (red dashed histogram). The limit of completeness of the sample is represented as a pointed line.}
\label{SampleA}
\end{figure}

The galaxies presented in this paper are a sub-sample of R08 which have available archival HST. The imaging is used to measure their gas radius and to derive their gas fraction. The targets have been gathered from the IMAGES-FORS2 survey, a representative sample drawn from both a spectroscopic and $M_J$ selection criteria (see R08). The final sample has 65 star-forming galaxies ($EW([OII])>15\,\AA\,$) and it is representative of star-forming galaxies at this redshift, see figure \ref{SampleA}. The objects with AGN contamination have been excluded from the final sample using diagnostic diagram and X-ray \& Radio observations, see detail in R08. We have tested the null hypothesis that the IMAGES-FORS2 and the final samples arise from the same parent population using a Kuiper test, a modified version of the two-tailed Kolmogorov-Smirnov test \citep{2010ApJ...712..318R}.  A standard confidence level of 0.05 or 95\% was selected for the rejection threshold. The Kuiper test results show that the distribution in K-band and stellar mass of our sample is indistinguishable from the IMAGES-FORS2 sample which itself was shown to be representative of intermediate-mass galaxies at z$\sim$0.6 (R08). 

\subsection{Data analysis}

The following section describes the methodology used to derived stellar mass, metallicities, SFR, gas radius and gas fractions.  These quantities and the associated errors are shown in Table 1 and 2.

\subsubsection*{Absolute magnitude}
The majority of the galaxies - 51 of 65 - belong to the Chandra Deep Field South (CDFS).  This field has been widely covered by UV to IR surveys including:  GALEX/NUV \citep{2007ApJS..173..682M}; HST/ACS (F435W [B], F606W [V], F775W [i], and F850LP [z] \citep{2006AJ....132.1729B};  EIS UÕ, U, B, V, R, and z bands in the optical \citep{2001A&A...379..740A}; GOODS-ISAAC $J$, $H$, and $K_S$ bands \citep{2010A&A...511A..50R} in the NIR.;  Spitzer/IRAC 3.6, 4.5, 5.8, and 8.0 $\mu m$ (Dickinson et al. in prep.).  We have obtained photometric measurements for galaxies in the sample by cross-correlating with catalogues in public archives or measuring the sources from data release images. All detections have been verified by eye and objects with possible flux contamination by neighbour sources have been flagged. The photometry was measured within an aperture of 3" diameter following \citet{2009A&A...493..899P}, and PSF correction in the UV and IR band have been applied. The rest of the sample are CFRS and UDSF galaxies for which optical, K-band and ISOCAM observation are available \citep{1999ApJ...517..148F}. 

\begin{figure}
\resizebox{\hsize}{!}{\includegraphics{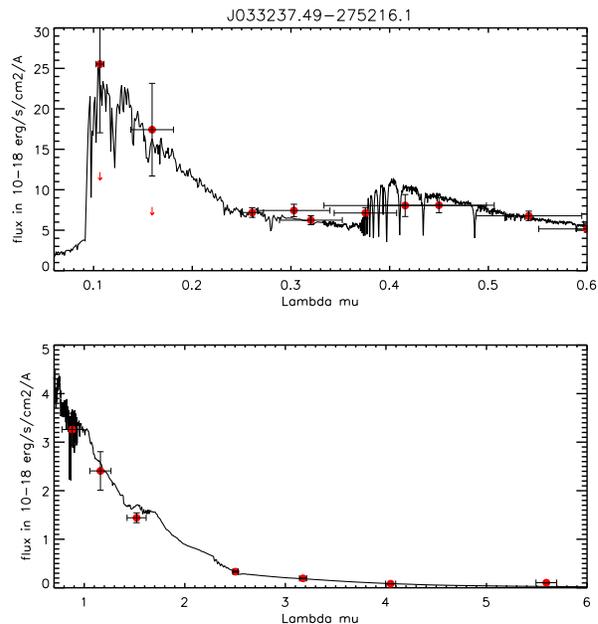}}
\caption{{\small Rest-frame SED of J033237.49-275216.1 for a 3 arcsec fixed aperture photometry (red symbols).  The SED has been modeled by a linear combination of 6 complex stellar populations  and a two component extinction law (solid line). The complex stellar population templates are parametrized by a exponential decline star formation rate with $\tau$=100Myrs and arise from Charlot \& Bruzual models. }}
\label{SED fitting}
\end{figure}

Absolute magnitudes in rest-frame have been measured using a K-correction approach. We have fitted the rest-frame SEDs with a linear combination of 6 stellar populations from \citet{2011ascl.soft04005B} models and a two component extinction law of \citet{1989ApJ...345..245C}, see Figure \ref{SED fitting}. Absolute magnitude have been computed from the best $\chi^2$ model. The fit gives low values of $\chi^2$ due to the large number of free parameters, thus this does not imply that the best solution is the real physical value. Indeed, the decomposition of the SED into stellar populations and dust extinction is a problem because these parameters are degenerate in particular age-metallicity and extinction. For this reason, we did not use the global properties derived from the SED fit such as extinction, SFR and stellar masses. However, this best-fit provides us with an accurate interpolation of the SED and thus enables us to derive absolute magnitudes with a high accuracy.

\subsubsection*{Stellar mass}

The stellar masses were derived from absolute magnitudes in K-band using the method proposed by \citet{2003ApJS..149..289B}. This method takes advantage of the tight correlation found between rest-frame optical colours and $M_*/L_K$ ratios, assuming a universal a diet Salpeter IMF. The $M_*/L_K$ ratios are corrected for the amount of light due to red-giant stars using $g-r$ colours. According to \citet{2003ApJS..149..289B}, the total random uncertainty on $log(M_*/M_\odot)$ at z$\sim$0 should be lower than $0.1\,$dex, and the systematic uncertainties due to galaxy ages, dust, or bursts of star-formation can reach 0.3\,dex. The stellar mass estimated at $z\sim0.6$ may suffer an additional systematic offset because the \citet{2003ApJS..149..289B} recipe does not take into account the evolution of $log(M_*/L_{K})$ with lookback time. \citet{2008A&A...484..173P} have estimated that the evolution of $log(M_*/L_{K})$ between z=0 and z=0.6 may be underestimated by up to -0.1dex. This result has been taken into account when comparing local and intermediate redshift samples.

We have been investigated the estimation of stellar mass in high-z samples using different SED methods (e.g. Single complex stellar population fitting by \citet{2004A&A...424...23F}, Maximum mass method by \citet{2001ApJ...559..620P}). The preliminary results are summarize hereafter, a paper is in preparation (Rodrigues et al. in prep). In the local Universe, a simple stellar population (e.g a SSP) or a SFH parametrized by a young and old component are enough to retrieve reliable estimation ofÊ global properties of most galaxies that form stars at moderate rates. On the contrary, at higher redshifts, most galaxies are in a phase of intense star formation, i.e., starbursts. Their photon budgets are dominated by the light emitted by young and massive stars in most of the spectral energy distribution which hide the old and intermediate stellar populations. The fraction of old stars is systematically underestimated by current SED fitting methods in distant starbursts \citep{2009ApJ...696..348W}. The stellar mass estimation derived by SED fitting is also dependent on the available photometric bands (for example if near IR rest bands are available). Using the same SED method on samples of different redshifts with different photometric band will not prevent strong systematics. For this reason we decide to use a conservative approach to derived the stellar mass in our sample and kept the \citet{2003ApJS..149..289B}  method which have been used in all the previous paper of the IMAGES survey.

\subsubsection*{Metallicity}

The procedure used to extract the gas metallicity was described in \citet{2008A&A...492..371R}. Briefly, we used high S/N -spectra, from the VLT/FORS2 spectrograph which has a moderate spectral resolution of $\Delta \lambda=6.8\AA$- to derive metallicities from the strong line ratio $R23$. The underlying stellar component has been subtracted from each individual spectra using the software Starlight \citep{2005MNRAS.358..363C} and the stellar library of \citet{1984ApJS...56..257J}. The extinction has been evaluated by two independent procedures: Balmer decrement and IR vs optical SFR balance. In order to compare the metallicities at $z\sim0.6$ and at $z\sim0$, we used the same metallicity calibration that the one use in the SDSS \citep{2004ApJ...613..898T}.  For the 2 galaxies the [OII] emission line falls out the wavelength range. We have used in this case the R3 parameter with the \citet{1992ApJ...401..543V} calibration. We have not found any evidence of systematic bias between the R23 and the R3 calibrations.\\

The $R23$ parameter is doubled-value with $12+\log(O/H)$, calibrations split into two branches: an upper-branch that corresponds to high metallicity values ($12 + \log{O/H} >8.4$)  and a lower-branch associated to low metal abundance. We have assumed that our sample of intermediate-mass galaxies is metal rich and metallicities have been estimated using the upper-branch calibration derived by \citet{2004ApJ...613..898T}. The validity of this assumption is confirmed by the values of the $O_3O_2$ indicator\footnote{The $O_3O_2$ parameter is defined as the ratio $[OIII]\lambda5007/[OII]\lambda3727$} \citep{2006A&A...459...85N}. All galaxies have $O_3O_2<$2  which place them into the metal-rich branch of the $R_{23}$ sequence. Moreover, galaxies with low values of $R23$ and low metallicities are extremely rare within out selected range of stellar mass. They are usually associated with dwarf galaxies. 

\subsubsection*{Star formation rate}
We have derived the SFR from three SFR calibrations: $H_\alpha$, $UV$ and $IR$. The three SFR rates have been calculated using the calibrations of \citet{1998ARA&A..36..189K} and a 'diet' Salpeter IMF. The associated error has been calculated by propagating the flux error through the SFR calibrations. A brief description of the methodology is given below, details can be found in R08. 

At the redshift range considered in this paper, the $H\alpha$ line is redshifted outside the optical window and we have estimated the $SFR_{H\alpha}$ from the $H\beta$ line. The $H\alpha$ flux is estimated by applying the theoretical ratio between $H\alpha$ and $H\beta$ \citep{1989agna.book.....O}. The $H\beta$ line have been corrected from dust extinction luminosity using the Balmer decrement and IR/$H\beta$ energy balance, see R08 for details. The aperture corrections have been determined from photometry in the available bands. For the CDFS targets, we derived the $SFR_{IR}$ from the $24\mu m$ flux from the DR3 MIPS catalogue \citep{2005ApJ...632..169L} and the \citet{2001ApJ...556..562C} relation. The  $SFR_{IR}$ of UDFS and CRFS target  have been inferred from the $15\mu$ ISOCAM observation using the calibration of \citet{2002A&A...384..848E}.  The $SFR_{UV}$ has been derived from the rest frame luminosity at 2800$\AA$ given by the SED fitting and the \citet{1998ARA&A..36..189K} UV calibration. The total recent star formation ($\sim100\,Myr$) has been estimated from the sum of $SFR_{IR}$ and $SFR_{UV}$. Indeed, SFR(UV) traces the UV light which has not been obscured by the dust and SFR(IR) traces the UV light thermally reprocessed by the dust. For the 24 objects without mid-IR detection, we defined the total SFR as being $SFR_{H\alpha}$ corrected by extinction, which corresponds to an instantaneous total SFR and is found by \citet{2004A&A...415..885F} to correlate well with $SFR_{IR}$ .

\subsubsection*{Gas fraction}

The mass of neutral and molecular gas can be measured using HI or CO observations. At present, the gas fraction of large samples of intermediate redshift galaxies can only be indirectly estimated by inverting the Kennicutt-Schmidt law. The method has been used successfully for local galaxies \citep{2004ApJ...613..898T}, z$\sim$0.6 \citep{2010A&A...510A..68P}, z$\sim$2 \citep{2006ApJ...644..813E} and z$\sim$3 \citep{2009MNRAS.398.1915M} galaxies. We have used the local K-S \citep{1998ARA&A..36..189K} with a power law indice of n=1.4 and assumed a 'diet' Salpeter IMF. Combining the \citet{1998ARA&A..36..189K} law and the definition of the SFR density, the gas mass is given by:
\begin{equation}
M_{gas}(M_{\odot})=5.188\times10^8\times SFR_{Total}^{\,0.714} \times R_{gas}^{\,0.572},
\label{eq_mgas}
\end{equation}
with $R_{gas}$ in $Kpc$ and $SFR_{Total}$ in $M_{\odot}\, yr^{-1}$. \\

We assume that the distribution of the ionised gas follows that of massive stars, which can be estimated using a rest-frame UV radius. In the case of a thin exponential disk, the optical radius is 1.9 times the UV half light radius $R_{UV}$ \citep{1991ApJ...368...60P} and $R_{gas}$ is then proportional to 1.9$\times R_{UV}$.  The rest-frame UV $R_{half}$ has been estimated from  B and/or V band HST images from the GOODs archive.  Archival HST ACS and WFPC2 observations from the UDSF and CRFS fields have been completely re-reduced.  This includes combining observations from multiple epochs to significantly increase the depth of the observations beyond those of the original individual programs (complete description of the re-analyse in Rodrigues et al. in prep). For each galaxy, we derived inclination, principal angle and $R_{half}$ following the procedure described in \citet{2008A&A...484..159N} and \citet{2010A&A...509A..78D}. The $R_{gas}$ error has been estimated by propagating the uncertainty of the $R_{UV}$ measurements. The uncertainties of $M_{gas}$ are the quadratic sum of the propagated error of $SFR_{total}$ and $R_{gas}$, and the uncertainty due to the dispersion of the Kennicutt-Schmidt law. The gas fraction is defined as:
\begin{equation}
f_{gas}=\frac{M_{gas}}{M_{gas}+M_*}
\end{equation}

There are several systematic effects which can affect the estimation of the gas fraction. First, we have assumed that the K-S law at $z\sim0.6$ follows the same power law than in local Universe. Some authors point out that the index of the K-S power law could change with redshift \citep{2007ApJ...671..303B} and thus induce systematic shifts in the estimation of the gas fraction at higher redshift. Nevertheless, the distributions of SFR surface densities of the samples are concentrated within a small range $\Sigma\,SFR=0.01-0.1\,M_\odot/yr/kpc^2$ which corresponds to an interval where the K-S relation is not affected by the variation of the power law \citep{2010A&A...510A..68P}. Therefore, a possible evolution of the power-law index of the K-S law, between z=0 and z$\sim$0.6, will only weakly affect the results. Second, in order to overcome the lack of direct measurement for the ionised gas distribution from integrated field spectroscopy, the distribution of young and massive stars from images in blue or UV bands have been frequently used in the literature as a proxy of the ionised gas. However, \citep{2010A&A...510A..68P} have shown that the extension of the ionised gas can be much more extended than the UV light in intermediate redshift galaxies, especially for compact galaxies. 
Using the GIRAFFE observations from \citep{2010A&A...510A..68P} and the deep imagery in B and V-band from HST/ACS, we have 
estimated a calibration between the $R_{gas}$ measured from the ionised extension and those estimated from the UV $R_{half}$.  
\begin{equation}
R_{gas} (Kpc)=0.91\times R_{UV} +  5.46 
\end{equation}
This correction is valid for compact system below $R_{UV}< 6.8\,kpc$. At larger $R_{UV}$, the measurement of $R_{OII}$ is underestimated since galaxies extend further than the GIRAFFE IFU field of view (3$\times$2 arcsec). When applying this correction to the data, the fraction of gas of gas-poor galaxies remains unchanged while those of gas-rich galaxies increase by $\sim5\%$. 

\section{Gaseous phase of intermediate redshift galaxies}
In this section we present the chemical properties of the gas of intermediate mass galaxies at $z\sim0.6$.

\subsection{State of the ionised gas}
\begin{figure} 
\centering
\resizebox{\hsize}{!}{\includegraphics{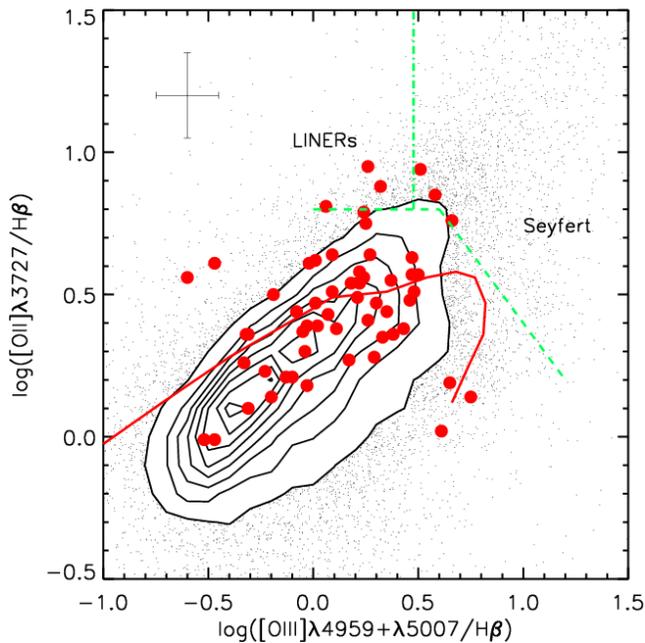}}
\caption{{\small  The [OIII]/$H\beta$ vs [OII]/$H\beta$ diagnostic diagram for the local SDSS galaxies (in contours and dots) and the intermediate redshift galaxies  (red circles). The solid line shows the theoretical sequence from \citet{1985ApJS...57....1M} for extra-galactic HII regions. The dashed line shows the photo-ionisation limit for a stellar temperature of 60\,000\,K and empirically delimits the Seyfert~2 and LINER from the HII regions. The dot-dashed line shows the demarcation between Seyfert~2 and LINERs from \citet{1989agna.book.....O}. The z$\sim$0.6 galaxies lie in the same locus in the diagnostic diagram than the local galaxies and they thus may share the same physical properties. }}
\label{SDSS_diag_OII}
\end{figure}

 \citet{2005ApJ...635.1006S} and \citet{2008ApJ...678..758L} have found that $z>1$ galaxies are offseted from the local relation in the diagnostic diagrams. They suggested that $z>1$ HII regions could have different physical properties than local counterparts: an harder ionising spectrum, higher ionisation parameter and more affected by shocks and AGN. We searched for evidence of different physical condition in the HII region of z$\sim$0.6 galaxies by comparing their positions with those of SDSS local galaxies in $[OII]\lambda3727/H\beta$ vs $[OIII]\lambda\lambda4959,5007/H\beta$ diagnostic diagrams. As shown in Figure \ref{SDSS_diag_OII}, the intermediate redshifts galaxies lie in the same locus as the SDSS galaxies. There is no evidence that HII region in intermediate redshift have different physical conditions than the local counterparts. This result is consistent with the previous work of \citet{1997ApJ...481...49H}, and for galaxies at lower redshift $0 < z < 0.3$, with \citet{1996MNRAS.281..847T}.

Four objects fall in the Seyfert and LINERs regions of the diagnostic diagram. These objects have not been excluded from the sample because they are compatible with HII regions when taking account their error bars. Moreover, X-ray detections by the Chandra X-Ray Observatory 4 Ms observations \citep{2011ApJS..195...10X} and radio detections from the Australia Telescope Compact Array 1.4 GHz \citep{2006AJ....131.1216A} have not been found for these objects.Ê

\subsection{Outflows}
Star-formation driven outflows may play a large role in the formation and evolution of galaxies and in enriching the intergalactic medium in metals \citep{2002ApJ...574..590S, 2006MNRAS.373.1265O}. In the local Universe, large scales gas outflows are linked to galaxies exhibiting high SFR densities \citep{1996ApJ...462..651L} such as starburst galaxies, LIRGs and ULIRGs \citep{2002ApJ...570..588R,2005ApJ...621..227M,2011ApJ...729L..27R}. As the density of the cosmic star formation rate increases with lookback time \citep{2004ApJ...615..209H}, outflows are expected to be more intense in the past. At high redshift, powerful large-scale winds, with velocities between $\sim200-2000 km/s$, have been detected in $z\sim0.6$ extremely luminous post-starburst \citep{2007ApJ...663L..77T}, post-starburst and AGN host galaxies at $0.2<z<0.8$ \citep{2011.Coil1104.0681}, low-ionisation outflows in $0.11<z<0.54$ star-forming galaxies \citep{2009ApJ...696..214S}, $z\sim1.3$ luminous starburst \citep{2011MNRAS..1571BB},  and star forming galaxies at $z\sim1.4$ \citep{2009ApJ...692..187W}. The outflows is typically detected in the UV by studying the kinematics of absorption lines from the interstellar medium, e.g. MgII, or by the morphology and kinematics of interstellar emission lines \citep[see review]{2004ASPC..320..277V}.   
 
\begin{figure} 
\centering
\resizebox{\hsize}{!}{\includegraphics{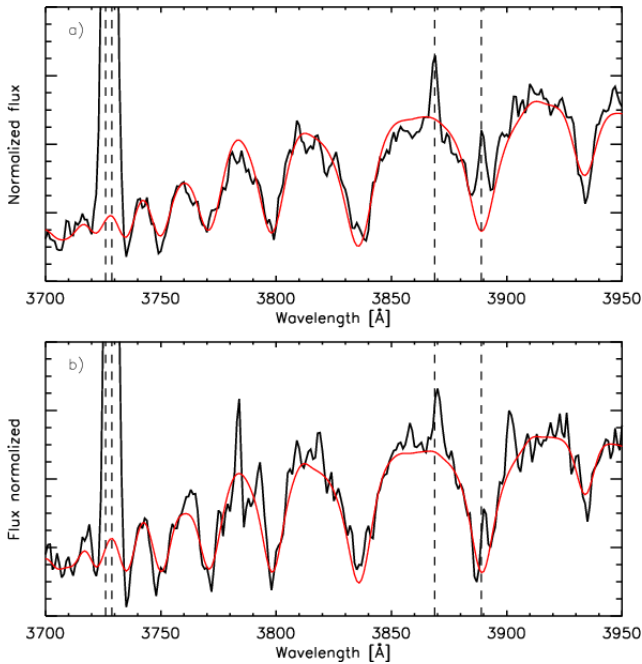}}
\caption{{\small Detection of  possible presence of outflows in intermediate mass galaxies at $z\sim0.6$ from FORS2 spectra (black solid line) from the difference of the velocity between gas emission lines and absorption lines produced by young stars. Synthetic absorption spectra estimated from STARLIGHT \citep{2005MNRAS.358..363C} are over-plotted with red lines. Spectra are rest-frame relatively to absorption lines. Dashed vertical lines indicate the position of the [OII]$\lambda\lambda3726.05,3728.8$, [NeIII]$\lambda3868.75$,  H8$\lambda3889.0$ lines in the rest position. The upper panel shows the stacked spectra of the 65 galaxies which constitute our sample. The fact that there is no shift between emission and absorption lines suggests that strong outflows are unlikely to be systematic process in progress in intermediate mass galaxies at this redshift. In the lower panel, we show an example of galaxies  J033225.26-274524.0 that probably host outflows. The observed lines are systematically shifted to the right relatively to the reference wavelength by $\sim162km/s\pm 75km/s$.    }}
\label{feedback_vel}
\end{figure}

We have first looked for the presence of outflows in the spectra of intermediate mass galaxies by measuring the difference of the velocity between gas emission lines and absorption lines produced by young stars (low order Balmer lines in the 3700\AA-4300\AA\,window), defined hereafter as $\Delta v$. At the given spectral resolution and taking into account the uncertainty on the wavelength calibration and on the determination of the position of the line, this method allow us to detected large scales outflows with velocities higher than $150km/s$. $\Delta v$ has been measured in the stacked spectrum of all targets in the sample and in individual spectra for a sub-sample of 15 galaxies having $S/N>10$ in the window continuum $4050\AA-4300\AA$. There is no velocity offset between emission and absorption lines in the stacked spectra which excludes the presence of systematic strong outflows in the population of intermediate mass galaxies, see Figure \ref{feedback_vel}. Systematic shifts of about $200km/s\pm 75km/s$ have been found in only three galaxies (J033224.60-274428.1, J033225.26- 274524.0, J033210.76- 274234.6). 
 
Next, we have investigated the morphologies of the $[OIII]\lambda5007\AA\,$ line. Galaxies hosting powerful outflows should show an asymmetric emission profile of their collisional lines with a tail in the blue wing, see Figure \ref{feedback_vel}. In the stacked spectra, the $[OIII]\lambda5007\AA\,$ line has a profile compatible with a gaussian distribution ($\chi^2\sim 1$). The line has a skewness of 0.60 and a kurtosis of -1.63. We search for a possible dependence of outflow on the stellar mass by stacking spectra into 6 stellar mass bins. The $[OIII]\lambda5007\AA\,$ line has a gaussian profile in all the mass bins and we thus conclude that there is no dependence on the stellar mass. Individual spectra also have $[OIII]\lambda5007\AA$ lines with gaussian shapes, except two galaxies: J033229.64-274242.6 and CFRS220599. Both galaxies have a two component profile of the [OIII] line compatible with the presence of a large scale outflow with $v_{wind}=445km/s\pm 50km/s$ and $v_{wind}=300km/s\pm 50km/s$ respectively. \\

To conclude, from the 65 galaxies in our representative sample of intermediate mass galaxies, only two hosts a powerful outflow with $v_{wind}=400km/s$  and three may have moderate outflows with $v_{wind}<300km/s$. This suggests that large-scale outflows do not play an important role in intermediate mass galaxies at $z\sim0.6$. The fraction of galaxies with outflows represents 8\% of the population of star-forming galaxies and 4\% of the population of intermediate mass galaxies at $z\sim0.6$ \footnote{Star-forming galaxies represent 60\% of the population of intermediate mass galaxies at this redshift, \citet{2010A&A...509A..78D}}. 

\begin{figure} 
\centering
\resizebox{\hsize}{!}{\includegraphics{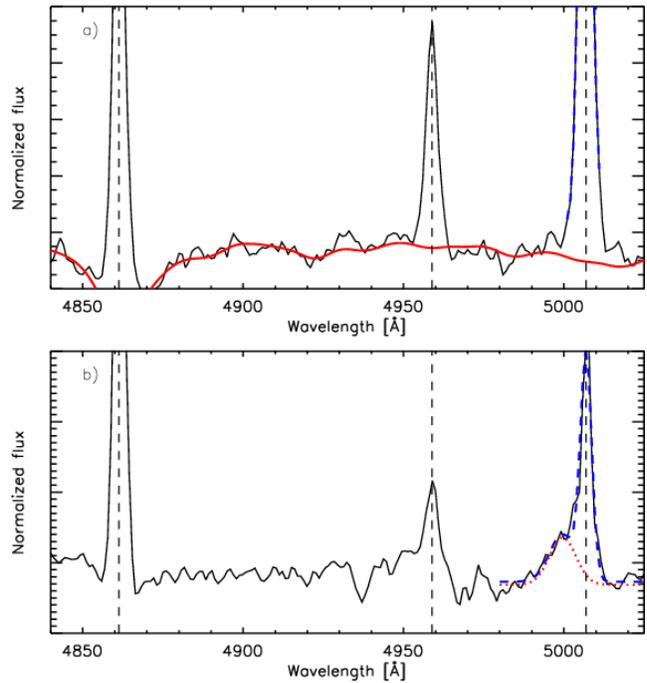}}
\caption{{\small Detection of  possible presence of outflows in intermediate mass galaxies at $z\sim0.6$ from FORS2 spectra from the morphology of the collisional line $[OIII]
\lambda5007$. Dashed vertical lines indicate the position of the $H\beta4861\AA$, $[OIII]\lambda4959\AA\,$ and $[OIII]\lambda5007\AA\,$ lines in the rest position. In the upper panel, we show the stacked spectra of the 65 galaxies which constitute our sample. The $[OIII]\lambda5007\AA\,$ line has a gaussian profile with a skewness equals to 0.66 and a S/N=120 (the blue line is a gaussian fit to the [OIII] line). In the lower panel we show the asymmetric profile of $[OIII]\lambda4959\AA\,$ line of J033229.64-274242.6 (S/N=32 and skewness=2.40), characteristic of outflows. This galaxy is the one of the two object in our sample that clearly host a strong outflow with $v_{wind}=445km/s$. Its $[OIII]\lambda5007\AA\,$ line is formed by two components: a narrow component with $FWHM=3\AA\,$ at $5006.9\AA\,$ and a broad component with $FWHM=13\AA\,$ at $4999.46\AA\,$. The blue-shifted broad component is over-plotted with a red dot line.}}
\label{feedback}
\end{figure}

\subsection{Gas fraction versus stellar mass at $z\sim0.6$}

We have completed the original sample of 65 intermediate mass galaxies with new observations of 25 galaxies from \citet{2010A&A...510A..68P}. These galaxies were observed using the VLT/GIRAFFE IFU instrument as part of the IMAGES survey. The objects were selected using the same criteria as those of this sample: $M_J>-20.3$ and $EW(OII)>15$\AA\, and both samples have homogeneous measurements of SFRs and $M_*$. The final sample (FORS2+GIRAFFE, hereafter gas~sample) is composed of 91 galaxies with a mean redshift of 0.666. We have verified the representativity of the sample by comparing the distribution of AB absolute magnitudes in K-band with the K-band luminosity function at z=0.5 and z=1 from \citet{2003A&A...402..837P}. The Kuiper test was applied to the $M_K$ distribution of the sample and those of the luminosity function, in the redshift range z=0.4 and z=0.98 and demonstrates that the Null Hypothesis cannot be rejected at better than the 95\% threshold. This means the two distributions arise from the same parent population.

\begin{figure} 
\centering
\resizebox{\hsize}{!}{\includegraphics{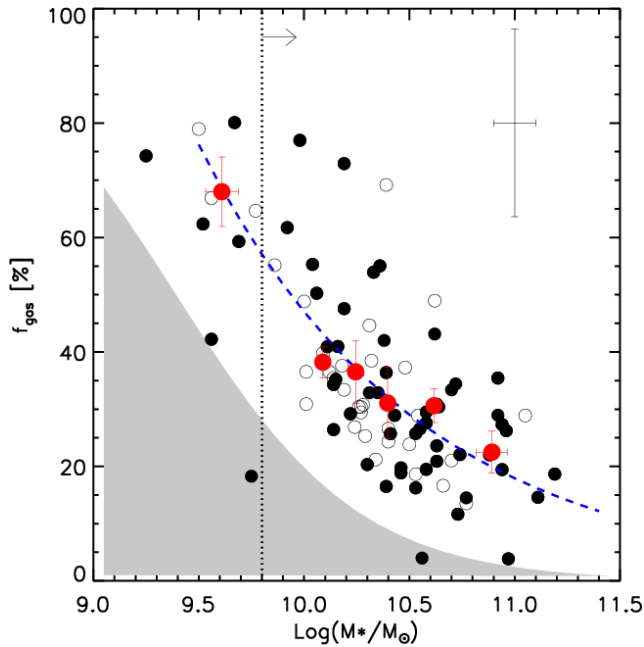}}
\caption{{\small The gas fraction derived from the inversion of the Kennicutt-Schmitt law as a function of the stellar mass, derived from the absolute magnitude in K band, for our sample galaxies (black symbols). Open symbols correspond to objects having only a lower limit of star formation rate. The red symbols correspond to the data binned in stellar mass. The grey area represents the region above the limit detection of the SFR on the gas fraction. The blue dashed line is the best $\chi^2$ fit of the data by an exponential function. The vertical black dotted line show the limit of completeness in stellar mass of the sample.}}
\label{GasFrac}
\end{figure}

Figure \ref{GasFrac} shows the gas fraction as a function of stellar mass for the 91 intermediate mass galaxies. It can be seen that these two quantities are strongly correlated, with lower stellar mass galaxies having higher gas fraction. This correlation has already been observed in the local Universe \citep{1997ApJ...481..689M,2000MNRAS.312..497B, 2004ApJ...611L..89K} and at higher redshift, e.g. \citet{2006ApJ...646..107E} at z$\sim$2 and \citet{2005A&A...433..807M} at z$\sim$3. We have fitted the data points with an exponential function and found that the best $\chi^2$ fit is the following:
\begin{equation}
\log{f_{gas}} [\%] = 5.86-0.41\times \log{M_*/M_\odot},
\label{equation_gas}
\end{equation}
with an 1$\sigma$ spread of the data about $\pm$ 12\%. Only galaxies having stellar masses over the limit of completeness have been taken into account in the fit. 

We have investigated if the distribution of $f_{gas}$ and $\log{M_*/M_\odot}$ was intrinsic or due to a limit on SFR detection. Indeed, the  sample is made up of emission lines galaxies with $EW([OII])>15$\AA\, and hence it is SFR-limited by construction. The limit in SFR induces a lower limit on the $M_{gas}$ which can be detected and thus affects the low mass region where only very gas-rich objects can be detected. The $f_{gas}$ detection limit has been evaluated using $SFR_{lim}$=2$M_\odot$/yr. The region above the $f_{gas}$ detection limit is indicated as a grey area in Figure \ref{GasFrac}. As the $z\sim$0.6 galaxies are well over the SFR limit threshold, we conclude that the shape of the $f_{gas}$ vs. $logM_*$ relation is real and that it is not due to any bias from the SFR limit.



\section{Evolution of gas fraction with lookback time}
In this section, we use gas estimates from the literature to extend the intermediate-redshift estimates from Sect 3 to larger lookback times and get a larger picture of gas evolution in intermediate-mass galaxies

\subsection{Comparing samples at different lookback times}
Our primary concern is whether samples of intermediate mass galaxies at different redshifts correspond to the same population of galaxies. In order to construct an evolutionary sequence, objects at different redshifts have to be linked together as progenitors and descendants. We have shown in a previous paper \citep{2010A&A...509A..78D} that the $M_J$ selected galaxies of the IMAGES survey are directly linked to local galaxies, and are the progenitors of local spirals. 
Above $z\sim2$, intermediate mass galaxies are no directly linked with their local counterparts. First, the higher merger rate at high redshift \citep{2009MNRAS.396.2345B} implies that most of galaxies could have experienced a major merger between z=2 and z=0. This prevents us from establishing a direct link between the two epochs. Moreover, simulations suggest that the local descendant of $z\sim2$ intermediate mass galaxy haloes are more massive, with $M_{halo}\sim4\times10^{12}M_\odot$ \citep{2009MNRAS.398.1858M}. A significant part of the population of intermediate mass galaxies at $z\sim2$ may have evolved into local elliptical galaxies. As a consequence, only the evolution of the properties of intermediate mass galaxies shown hereafter can be seen as an evolutionary sequence of local spiral galaxies between z=0 to z=0.6, while at higher redshift we can describe only the evolution of the gas fraction and metallicity of galaxies within a stellar mass range at a given epoch.
 
A second concern is whether the observed samples are representative of all of the populations of intermediate mass galaxies at a given epoch. The $z\sim0.6$ sample has been selected by absolute luminosity in restframe J-band, which is a good proxy for the stellar mass \citep{2010A&A...509A..78D}.  However, since the aim of this work is to study the properties of the ionised ISM, only star-forming galaxies with emission lines $EW([OII])>15$\AA\, have been selected. Our samples do not include quiescent galaxies which represent 40 \% of galaxies at z=0.6 and more than 80\% in the local Universe \citep{1997ApJ...481...49H}. However, at intermediate redshift, quiescent galaxies are expected to be massive gas poor and quiescent from $z=0.6$ to the present. Thus this population of galaxies has a limited contribution to the gas and metals evolution. On the other hand, an important bias may arise from the population of Low Surface Brightness galaxies (LSB) which are not included in our sample. Studies in the local Universe have shown that LSBs have low gas mass density, small SFR and low metallicity but have surprisingly high gas fractions (more than 50\% of their baryonic mass). Despite their high gas content, the gas is rather extended in LSBs and their gas densities are typically below the star-formation threshold of the Kennicutt-Schmidt law. This explains why LSBs have low SFRs. At z=0, LSBs represent around a few percent \citep{1997AJ....114..635D} of the galaxy population. However the evolution of their number density with redshift remains unknown. A strong evolution of LSBs could have important consequences on the models of galaxy evolution since this population may provide a large neutral gas reservoir. 
 
\subsection{Comparing samples with different gas mass tracers}
The gas fraction were estimated using different methods as a function of redshift: In the local Universe, the gas mass can be measured directly from the HI flux; At high redshift, gas fraction are mainly indirect measurements estimated using the inverse K-S law. In the last year, CO measurements have been made for a limited sample of galaxies up to $z\sim2.2$. When comparing gas fraction at different redshifts, it is important to keep in mind these estimations of the gas mass may not trace exactly the same radius of gas. For instance, HI measurements take into account the entire cold gas of a galaxy while the estimation of the gas fraction via the K-S law traces the star-formation in the optical radius. Therefore it traces more closely the molecular gas which can collapse and form stars. The HI gas fraction is an upper limit of the total amount of gas. At higher redshift, the CO measurements and the K-S law both trace the dense molecular gas and therefore both methods give lower limits for the total gas fraction.

\subsection{Gas fraction and metallicity samples in the literature}
We have compared the gas fractions of intermediate redshift galaxies with those of published samples at different redshifts, see figure \ref{Fgas_evo}. In order to compare homogenous gas fractions and stellar masses at different redshifts, all of the gas fractions and stellar masses have been converted to the same Salpeter 'diet' IMF \citep{2003ApJS..149..289B}. Hereafter, we describe the different samples and discuss about the results.

\subsubsection*{Local direct HI mass samples}
 In the last few years, new large HI surveys, such as Arecibo Legacy Fast ALFA (ALFALFA) and the GALEX Arecibo SDSS survey (GASS) have produced a comprehensive studies of the gas content of local galaxies. Unfortunately, the depths of blind HI surveys are too shallow compared to volumed limited optical surveys such as SDSS. The issue of representativity of these HI samples must be considered with caution when comparing to optically selected samples. 
In Figure \ref{Fgas_evo} we have plotted  the trend of HI mass fraction as a function of the stellar mass for two local samples with direct HI measurements: a sample of UV-selected star-forming galaxies from \citet{2008AIPC.1035..180S} and the mass selected sample of \citet{2010MNRAS.403..683C} from the GASS survey. 

\subsubsection*{ $z\sim2$ sample: \citet{2006ApJ...647..128E}}
The \citet{2006ApJ...647..128E} galaxies are drawn from the rest-frame UV selected $z\sim2$ spectroscopic population described by \citet{2004ApJ...604..534S}. \citet{2006ApJ...647..128E} observed a sub-sample of these UV-selected galaxies.  This sub-sample was chosen based on properties such as morphology or IR magnitude. We have used this near-IR sub-sample as a comparison sample.
In order to homogenise measurements and methodologies between the $z\sim2$ and $z\sim0.6$ samples, we have re-derived the gas fractions of \citet{2006ApJ...644..813E} from the tabulated $SFR_{UV}$, extinction corrected $SFR_{H\alpha}$, $R_{gas}$ and $log\,M_*$ in \citet{2006ApJ...644..813E,2006ApJ...647..128E,2006ApJ...646..107E}. The gas extension has been estimated by assuming they are equal to $FWHM^2_{H\alpha}$ in \citet{2006ApJ...644..813E} where $FWHM_{H\alpha}$ is the extension along the slit of the $H\alpha$ emission in the spatial direction, while the gas extension in IMAGES sample have been estimated assuming a circular distribution of light equal to $\pi\,R^2_{gas}$. Following \citet{2010MNRAS.tmp..689P}, we have assumed the $R_{gas}$ in \citet{2006ApJ...646..107E} to be $2\sigma$ of the $H\alpha$ emission distribution and that the gas surface distribution  is Gaussian. $R_{gas}$ is then equal to $(2/2.35)\,FWHM_{H\alpha}$. This new derivation leads to slightly higher gas fraction by about +5\%. The median gas fraction of the sample is 53\%$\pm 5$ for a median $log\,M_*$=10.41, compared to 50\% as derived by \citep{2006ApJ...644..813E}. 

It is important to note that this sample may suffer of selection effect inherent to high-z studies and it may not be representative of star-forming galaxies at $z\sim2$. It is difficult to determine if the UV selected galaxies at $z\sim2$ are truly representative of all star-forming systems at this redshift.  The criteria of selection may preferentially select red and bright objets and therefore mainly massive galaxies at these redshift (see figure 5 of \citet{2006ApJ...646..107E}).

\subsubsection*{$z\sim3$ sample:  \citet{2009MNRAS.398.1915M}}
In \citet{2009MNRAS.398.1915M}, the authors present preliminary results for the sizes, SFRs, morphologies, gas-phase metallicities, gas fractions and effective yields of a sample of z$\sim$3 Lyman-break galaxies. However, they have not considered companions and secondary peaks of emission when measuring the gas extension. Therefore the gas mass is only consistently derived for the central region and the gas fraction estimation may suffer of strong underestimation. We have thus decided to keep this sample in the comparison as lower limits. 

\subsubsection*{High-z samples from CO estimation}
\begin{figure*}
\centering
 \includegraphics[width=8.5cm]{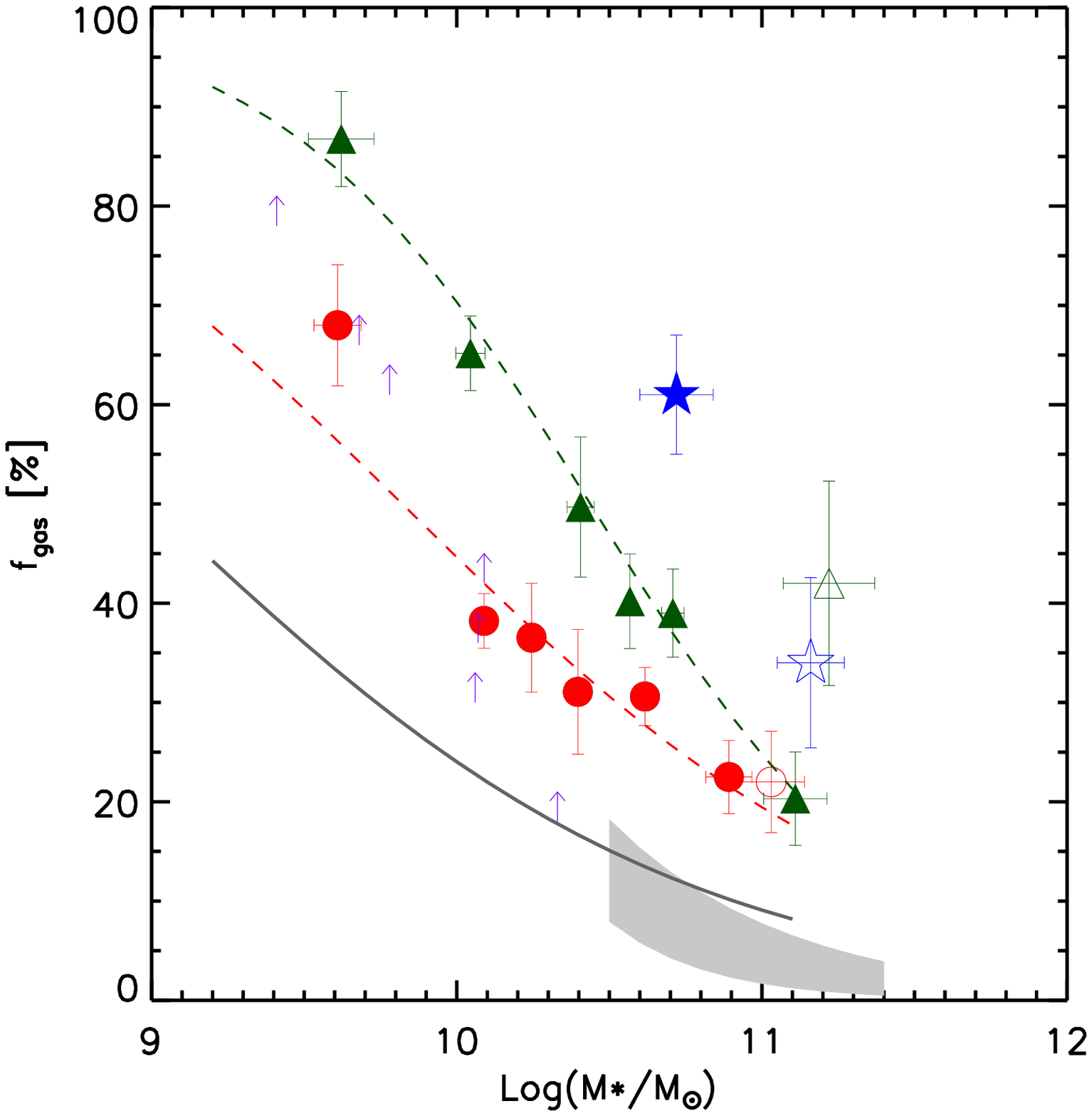}
  \includegraphics[width=8.5cm]{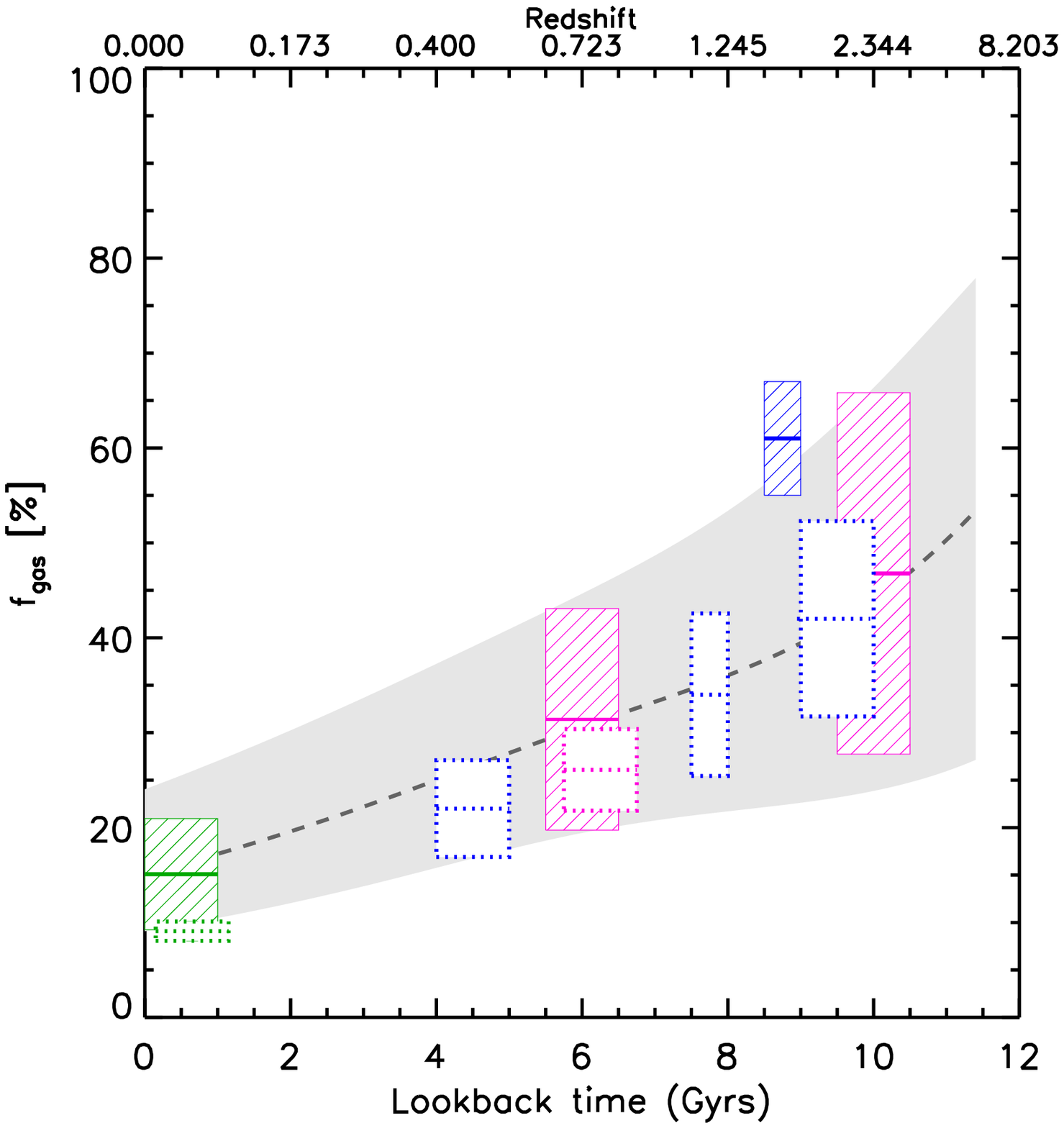}

\caption{{\small \textbf{Left}: The gas fraction $f_{gas}$ vs. stellar mass in local, intermediate and high-z samples. Two local galaxies samples are plotted as reference in grey. The \citet{2008AIPC.1035..180S}  sample composed by UV-selected star-forming galaxies is represented as a solid line. The data from the GASS survey is represented as a grey region. The upper limit corresponds to the median of the sample where the HI mass of none-detection have not been taken into account, while the lower limit is the median of the sample with HI mass of none-detection set to a upper limit. For clarity the $z\sim0.6$ (red dots) and $z\sim2$ from \citet{2006ApJ...647..128E}(green triangles) samples have been represented in bins of $\sim$12 and $\sim$15 galaxies respectively. Each point represents the median gas fraction and stellar mass in the bin. The errors bars correspond to the error on the evaluation of the median value calculated by bootstrap resampling. The direct measurement of the gas fraction at $z\sim0.4$ for a sample of 4 LIRGs from \citet{2011arXiv1102.3694G} is plotted as a red open circle. The gas fraction at $z\sim1.5$ from the CO observation of \citet{2010ApJ...713..686D} is plotted with a blue star. The direct measurements of CO gas mass from \citet{2010Natur.463..781T} are plotted as open blue star for the $z\sim1$ sample (9 objects) and as green triangle for $z\sim2$ (10 objects). The $z\sim3$ sample of \citet{2009MNRAS.398.1915M} is represented as lower limit arrows. \textbf{Right}: The gas fraction $f_{gas}$ as a function of lookback time for the sample described in section 4.2. The boxes with diagonal dashed correspond to intermediate mass samples, having stellar mass between $9.8<\log{M*/M\odot}<11.5$. The dotted line boxes are samples with stellar mass superior to $\log{M*/M\odot}=10.8$. The colours code how gas fractions have been estimated: green for HI measurements, pink for inverse K-S law estimations and blue for CO estimations.  The parametrization of $f_{gas}$ as a function of lookback time described by equation \ref{parametrization} for three stellar mass is represented in grey: $\log{M*/M\odot}=10.5$ in dashed line, $\log{M*/M\odot}=10$ upper limit of the grey area, and $\log{M*/M\odot}=11$ lower limit of the grey area. }}
\label{Fgas_evo}
\end{figure*}

We describe here  other samples from the literature which are shown in figure \ref{Fgas_evo}, but that have not been used in the estimation of the evolution of the gas fraction due to their incompleteness. 

\citet{2011arXiv1102.3694G} have recently measured the cold gas content from CO ($J=1\rightarrow0$) detection in a sample of $z\sim0.4$ galaxies. The sample is composed of seven LIRGs galaxies with $\langle\log{M_*/M_\odot}\rangle\sim11$. This sample has the same characteristics as galaxies in the higher mass bin of our sample therefore gas fractions can then be directly compared with ours. The gas fractions from CO measurement agree remarkably well with our measurements from the K-S inversion (see Figure \ref{Fgas_evo}).

In \citet{2010ApJ...713..686D}, the authors have estimated the gas fractions of normal, near-IR selected galaxies at z$\sim$1.5 using CO observations and simulations. They have used dynamical models of clumpy disk galaxies to derive dynamical masses and then estimate the gas masses. To be consistent with the gas fractions derived in the $z\sim0.6$ sample, we have re-estimated the gas fractions in the \citet{2010ApJ...713..686D} sample using the inverted K-S law and the tabulated values of gas radius $R_{CO}$, SFR and $logM_*$. This new derivation leads to a median gas fraction of 61\%$\pm 6$ for a median $log\,M_*$=10.72. This value is in very good agreement with the gas fraction derived by \citet{2010ApJ...713..686D} from the simulation. \citet{2010Natur.463..781T} present the results of a survey of molecular gas in samples of typical massive-star-forming galaxies at mean redshifts $\langle z\rangle$ of about 1.2 and 2.3. The typical stellar mass of the galaxies in this samples are $10^{11}M_\odot$.

\subsection{Evolution of the gas fraction}
Figure \,\ref{Fgas_evo} shows the evolution of the $f_{gas}$ vs. $log\,M_*$ relation and the evolution of the mean $f_{gas}$ as a function of  lookback time. The $z\sim0.6$ galaxies have a gas fraction twice that of their local counterparts \citep{2008AIPC.1035..180S} at $log\,M_*=10.43\,M_\odot$ . For reference, the  Milky Way has a $f_{gas}$=12\% \citep{2006MNRAS.372.1149F} and M 31 has $f_{gas}$=5\% \citep{2006ApJ...641L.109C}. At intermediate redshift, $0.4<z<0.7$, the estimation of the gas fraction from the inverse K-S and from CO measurements are in good agreement: CO measurements at $z\sim0.4$ give $f_{gas}=$ 20\%  and inverse K-S law estimation give $f_{gas}=24\%$ at $z\sim0.6$ for massive galaxies.  

At higher redshifts the quantitative evolution of the gas fraction is more complicated to constrain because of the inhomogeneity of the samples. For instance, the gas fractions estimated from the inversion of the K-S law at $z\sim2$ by \citet{2006ApJ...644..813E} are lower than those derived at $z\sim1$ and $z\sim2$ by CO observations for all of the stellar mass range. This discrepancy may arise from the fact that the CO IR-selected samples and the UV-selected sample of \citet{2006ApJ...644..813E} do not sample the same galaxy population. The \citet{2006ApJ...644..813E} selection may preferentially select gas poor galaxies compared to the other IR-selected samples. Taking into account only the CO observations, the mean gas fraction of intermediate mass galaxies reaches 60\% at $z\sim1.5$ and high mass galaxies increases their mean gas fractions from 35\% at $z\sim1$ to $40\%$ at $z\sim2$. The mean gas fraction increases linearly from $t_{lb}=0$\,Gyr to $t_{lb}=12$\,Gyr  at the rate of 3\% per Gyr for $\log{M*/M\odot}>10.8$ galaxies and per 4\% by Gyr for intermediate mass galaxies. 

We have parametrized the evolution of the gas fraction as a function of the stellar mass and lookback time using a similar formalism to that of \citet{2009MNRAS.397..802H}. The relation of \citet{2008AIPC.1035..180S}  at z=0, the IMAGES sample at $z\sim0.6$ and the \citet{2006ApJ...644..813E} sample at $z\sim2.2$ have been fitted using the following function:
\begin{equation}
f_{gas}(M_*| t)=\frac{1}{1+ (M_*/10^{A(t)})^{B(t)}} 
\label{parametrization}
\end{equation}
where t is the lookback time in Gyr, see Fig \ref{Fgas_evo}.
The evolution of the relation as a function of time is parametrized by the two functions, A(t) and B(t) given by: 
\begin{eqnarray}
A(t)&=&(9.00\pm0.06) + (0.13\pm0.017) \times t \\
B(t)&=&0.5+(13.36\pm68.22)\times e^{-(38.02\pm53)/t} 
\end{eqnarray}
A(t) represents the stellar mass at a given time for which the gas fraction is equal to 50\%. The parameter increases linearly with lookback time. It implies that the gas fraction in the population of intermediate mass galaxies increases on average linearly. On the other hand, $B$ corresponds to the slope of the function. Its exponent increases as a function of time suggesting that the evolution of the gas content is faster in high mass galaxies at higher redshift. However, this parameter is very sensitive to selection bias and the change of slope at higher redshift must be taken with caution. 


\section{Chemical evolution of intermediate mass galaxies in past 6 Gyr}

In this section, we compare the observed quantities with the prediction of a simple chemical model.  As a zero-order approximation of chemical evolution, the closed box model \citep{1981ARA&A..19...77P} provides a first insight into the metal enrichment of galaxies. 
In a simple closed-box model, galaxies are assumed to be a closed system without interaction with their environment, homogenous and well mixed boxes. Galaxies evolve passively, transforming their initial mass of pristine gas $M_{Gas}$ into stars $M_{*}$ and enriching the ISM by newly formed metals $M_Z$. The chemical cycle of galaxies can be described by the following system of equations \citep{1972ApJ...173...25S,1979MNRAS.189...95P}:
\begin{eqnarray}
 \frac{dM_{Gas}}{dt}&=&-\Psi(1-f_{recy}) \\
 \frac{dM_{*}}{dt} &=&\Psi(1-f_{recy}) \\
 \frac{dM_{Z}}{dt}&=&\Psi(Z_g+Z_{eject}\times f_{recy}) 
 \label{eq_chemical_ev}
\end{eqnarray}
where $\Psi$ is the rate of star formation, $f_{recy}$ and $Z_{eject}$ are respectively the rate to which gas is returned to the ISM and the metallicity ejected by the stars into the ISM after their death, $Z_g$ is the metallicity of the gas. The analytical integration leads to a relation between the metallicity of the gas and the gas fraction:  
\begin{equation}
Z_g(t)=Z^{ini}-y\times ln\frac{f_{gas}^{ini}}{f_{gas}(t)}
\label{eq_chemical_solution}
\end{equation}
where $y$ is the yield, defined as the rate at which metals are being returned to the ISM relatively to the current SFR, and $Z^{ini}$ and $f_{gas}^{ini}$ are the initial metallicity and gas fraction of the galaxies when starting to evolve as a closed box. 

\subsection{Metallicity versus Gas fraction at z$\sim$0.6}

\subsubsection*{Closed-box model}
\begin{figure} 
\centering
\resizebox{\hsize}{!}{\includegraphics{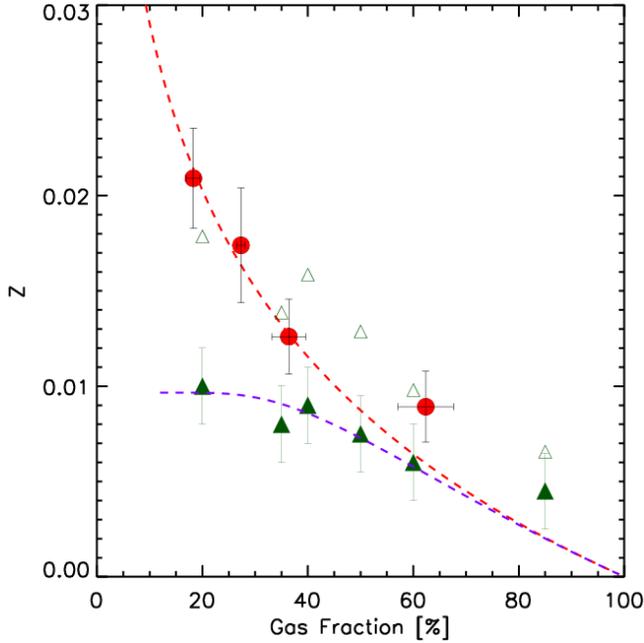}}
\caption{{\small  Evolution of the gas fraction - metallicity relation in the past 6 Gyr. The red circles correspond to the $z\sim0.6$ sample and the fill green triangle to the $z\sim2.2$ sample from \citet{2008ApJ...674..151E}. The open green triangles correspond to the \citet{2008ApJ...674..151E} sample with metallicity re-scale into R23 Tremonti calibration using \citet{2008arXiv0801.1849K} relations. The red dashed line is the theoretical relation in the case of a closed-box model with a yield equal to solar. The green dashed line is the best fit of the \citet{2008ApJ...674..151E} objects to models with infall and outflow: $f_{inf}=1.3$, $f_{outflow}=1.6$. }}
\label{Yield_evolution}
\end{figure}

At a given epoch a population of galaxies in a closed-box regime should verify equation \ref{eq_chemical_solution}. In figure \ref{Yield_evolution}, we have plotted the metallicity versus the gas fraction for galaxies of our sample in bin of gas fraction and compared them with the prediction of the closed-box model assuming a nucleosynthetic yield equals to solar $y=y_\odot=0.0126$ \citep{2004A&A...417..751A}. The data follow closely the prediction of the closed-box and suggests that the population behaves like a closed system. Notice that the closed-box model track is not a best-fit to the data points but just the representation of the equation \ref{eq_chemical_solution} assuming the usual nucleosynthetic yield in the literature. The uncertainties on the value of the yield are negligible in this case since the best-fit of equation \ref{eq_chemical_solution} to the data with the yield as a free parameter doesn't drastically improve the quality of the fit and gives a yield close of $0.92y_\odot$ (This will be discussed in more detail in the next section) . \\
 
\subsubsection*{Models with infall and outflows}
Recently, severals studies of the chemical evolution of galaxies have introduced  infall and outflow of gas into analytical chemical evolution models (equation 8-10) and they have constrained the contribution of these processes at different lookback times \citep{2006ApJ...647..128E,2009MNRAS.398.1915M,2008ApJ...674..151E}. These kinds of studies need to assume hypothesis on the properties of the infalls and outflows. Nevertheless, these toy models are interesting tools since they give qualitative descriptions of the chemical evolution of galaxies. We have investigated the upper limits of the contributions of infall and outflow in the population of intermediate mass galaxies by comparing our data with the model proposed by \citet{2008ApJ...674..151E}. This model can be used to reproduce $Z$ and $f_{gas}$ by changing three parameters, the amount of infall $f_{i}$ and outflow $f_{o}$ as a fraction of SFR and the yield $y$: 
\begin{equation}
Z=\frac{y\alpha}{f_i}\times(1-\frac{f_{gas}}{1+(1-f_{gas})(f_o/\alpha-f_i/\alpha)}^{\frac{f_i}{\alpha-f_i+f_o}} )
\end{equation}
where $\alpha$ is the fraction of mass remaining locked in stars. We assumed $ \alpha=1$ \citep{2008ApJ...674..151E}  i.e. the gas returned to the ISM by star formation is neglected \footnote{Notice that this assumption may maximize the contribution of infalls and outflows. }. This model assumes the following hypothesis: (1) the metallicity of the inflowing gas is negligible compared to that of the galaxy; (2) the outflowing gas has the same metallicity as the gas that remains in the galaxy; (3) the gas returned to the ISM by star formation is negligible. 

We have found that only models with a small fraction of infall, $f_{i}\sim0.23\times SFR$, and no outflow are consistent with $z\sim0.6$ galaxies. Given the mean SFR of the sample $<SFR>=11\,M_\odot/yr$ and assuming a solar yield, the contribution of infall to the population of intermediate mass galaxies only represents $4\pm2 M_\odot/yr$ per galaxy and the contribution of outflow between $1\pm 1 M_\odot/yr$.

\subsubsection*{Uncertainties from the assumed yield}
This results depends on the value of the assumed nucleosynthetic yield. In this work, we have assumed a solar yield $y_\odot$. However, the absolute value of the yield is only known within large uncertainties, with possible values spanning the interval $0.6y_\odot<y<1.8y_\odot$ \citep{2008MNRAS.385.2181F}. This large uncertainty results from the poor constraints on the initial mass function and the value of the SN II nucleosynthesis yield. Despite the fact that the absolute value is unknown, comparable studies can be done, even at different redshifts, because the yield is expected to remains constant up to high redshift. The yield can be expressed as the following:
\begin{equation}
y=\frac{dM_{new\,Z}/dt}{dM_{recy}/dt}\times \frac{dM_{recy}/dt}{\Psi}=Z_{eject}\times f_{recy} 
\label{eq_yield_true}
\end{equation}
where $dM_{new\,Z}$ is the rate of metal production from Type II supernovae, $dM_{recy}$ is the mass of gas return to the ISM after stars death. Thus the yield depends on the star formation history, the metallicity of the gas and the initial mass function. However, \citet{2010arXiv1007.3743P} have shown that the yield only depends on the star formation history during the first $10^7yr$, i.e the lifetime of stars that produce oxygen, and that it can be considered as a constant at longer timescales. Besides extreme cases of bursty and violent star formation histories, the yield remains constant with time and has small dependence on the metallicity of the gas. \\

\subsection{Metallicity versus Gas fraction at higher redshift}

Figure \ref{Yield_evolution} shows the evolution of the $f_{gas}$-Z relation during the last 6 Gyr. The fraction of the gas infall and outflow contribution in the sample of $z\sim2.2$ galaxies have been re-evaluated assuming the same yield as lower redshift studies. We found that the exchange of gas with the environment of the population of $z\sim2.2$ galaxies is high with $f_{infall}=1.3\times SFR$ and $f_{outflow}=1.6\times SFR$ which corresponds to $f_{infall}=78M_\odot\,yr^{-1}$ and $f_{outflow}=96M_\odot\,yr^{-1}$. This is lower than the previous estimate from \citet{2008ApJ...674..151E} where the yield was set as a free parameter of the fit. In fact, the fraction of $f_{outflow}$ and $f_{infall}$ strongly depends on the assumed yield. When the yield is set as varying inside the possible values of the nucleosynthesis yield, the fraction of $f_{outflow}$ and $f_{infall}$ are systematically higher. Such an effect is due to the presence of relatively metal-rich galaxies with high gas fraction that force the models to converge to high values of the yield. Consequently it adjusts higher values of $f_{outflow}$ and $f_{infall}$ in order to fit the gas-poor galaxies with effective yields close to a solar one. As mentioned in the previous section, there is no evidence for a change of the nucleosynthetic yield with time and that an variation of the yield at higher redshift will trend towards lower values. This new estimation does not change drastically the conclusion of \citet{2008ApJ...674..151E} and \citet{2009MNRAS.398.1915M} who claimed that high-z galaxies are very far from a closed-box model. The estimated high fraction of infall and outflow at high redshift ($\sim$1.5 times the SFR) supports the claim that accretion of gas at approximately the gas processing rate is the main driver of distant galaxies evolution beyond $z\sim2$. 

The estimation of the contribution of outflows and infall using the shape of the $f_{gas}$-Z relation is very sensitive to the metallicity calibration used\footnote{\citet{2008arXiv0801.1849K} have shown that the shape of the M-Z relation is very sensitive to the used calibration} (see discussion on the systematics from the metallicity calibration in R08). Converting the metallicities of the $z\sim2$ sample estimated with the N2 parameter to the Tremonti calibration \citep{2008arXiv0801.1849K} will push the point of low gas fraction to higher metallicities and the amount of outflow and infall estimated by the chemical models would drastically decrease (close to a closed-box). Unfortunately, the spectra of our sample and those of  high-z samples do not have enough spectral coverage to estimate the metallicities using different empirical calibrations to check that any result is not caused by the metallicity calibrations. Until metallicities have not been  estimated by several calibrations, the results of chemical models with infall and outflows have to be taken carefully and only qualitatively.

\subsection{Co-evolution of gas fraction and metals in the past 6 Gyr}

In the previous section, we have compared two relatively simply toy-models for describing the gas fraction and metals.  This comparison indicates that the observations are more consistent with a closed-box model than that of infall/outflow. This result does not imply that all intermediate mass galaxies are closed systems without any interaction with the IGM or other galaxies. Individual galaxies may have experienced interactions and mergers with each other. However, statistically the intermediate mass galaxies as a population can be described as closed-box systems. In this section, we aim to test if the observed metallicities and gas content of local galaxies are consistent with a scenario in which $z\sim0.6$ galaxies have evolved in isolation until now. \\ 

\begin{figure} 
\centering
\resizebox{\hsize}{!}{\includegraphics{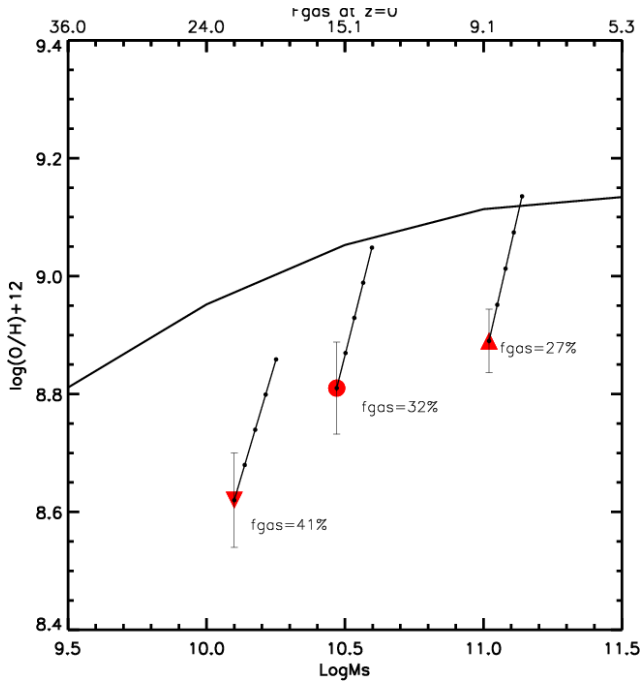}}
\caption{{\small  M-Z relation for $z\sim0.6$ galaxies (symbols) and the local relation from \citet{2004ApJ...613..898T} (solid line). The track of an evolution in closed-box is represented as thin solid line, each step of the track corresponds to a variation of 5\% in gas fraction. The $z\sim0.6$ galaxies have been divided into quartiles: the first quartile is the down-facing triangle ($\log{M_*/M_\odot}<10.2$), 2nd and 3rd quartile are the filled red circle ($10.2<\log{M_*/M_\odot}<10.8$) and the last quartil is the upward facing red triangle($\log{M_*/M_\odot}>10.8$). The median gas fraction in each bin is labeled close to each point. As reference, the gas fraction of local galaxies at a given stellar mass from \citet{2008AIPC.1035..180S} is plotted in the top axis.
}}
\label{M-Z_closebox}
\end{figure}

In the closed-box model, the variation of the metallicity as a function of the variation of gas fraction can be determined by deriving equation (\ref{eq_chemical_solution}) \citep{2006A&A...447..113L}. If we assume that the yield is constant with time, the derivative of equation \ref{eq_chemical_solution} as a function of gas fraction is given by: 
\begin{equation}
\mathrm{d}(\log{Z})/\mathrm{d}f_{gas}=\frac{0.434}{f_{gas}\ln{f_{gas}}} 
\label{eq_metalsevol}
\end{equation}
In parallel the stellar mass increases as: 
\begin{equation}
\mathrm{d}(\log{M*})/\mathrm{d}f_{gas}=-\frac{0.434}{1-f_{gas}} 
\label{eq_starsevol}
\end{equation}
Equation (\ref{eq_metalsevol}) implies that the increase of the metallicity in a closed box model is independent of the yield and only dependent on the variation of the gas fraction.
In Figure \ref{M-Z_closebox}, we have plotted the M-Z relation for galaxies in our sample and the local relation from \citet{2004ApJ...613..898T}. Using equation (\ref{eq_metalsevol}) and (\ref{eq_starsevol}) we have predicted the track in the M-Z plot of $z\sim0.6$ galaxies in three stellar mass bins for isolated evolution. Each step of the track corresponds to a variation of 5\% in gas fraction. After exhausting 20\% of their gas, $z\sim0.6$ galaxies reach the local M-Z relation. Their final gas fractions are equal to those of their local Universe counterparts at given stellar mass. The observed co-evolution of the gas fraction and metallicity from $z\sim0.6$ to $z=0$ is thus compatible within the error bars with a closed-box evolution. This later results supersedes our conclusion from R08, where we estimated that $\sim$30\% of the stellar mass of local galaxies may have formed through an external supply of gas. The strong constrain given by the evolution of the gas fraction derived in this paper allow us to lead now to a more robust result. 

For the low mass bin the conclusion is less clear. Figure \ref{M-Z_closebox} shows that the low mass bin does not reach local metallicities after exhausting the observed variation of gas fraction. This implies that low mass galaxies ($\log{M_*/M_\odot}<10.2$) may not follow an isolated evolution. This effect may also be seen in the $f_{gas}-Z$ relation at $z\sim0.6$, where the bin containing the low-stellar mass galaxies with high gas-fractions is offset from the prediction of the closed-box mode. Unfortunately the low mass bin is more likely to be affected by systematics, such as metallicity estimation ( values close to turnover of the $R23-12+\log{O/H}$ relation) and selection effect (preferential selection of gas-rich objects). Based on this we cannot confidently state that this trend is real.  \\

\section{Discussion and conclusion}
Our observations suggest that the population of intermediate mass galaxies have evolved as closed systems over the past 6\,Gyr, in contrast with its past episodes of intense outflow and infall  at $z>2$. This result only implies that the observed evolution of this population is compatible with a closed-box but it cannot exclude possible contribution of infalls and outflows on a case-by-case basis. An evolution in isolation from $z\sim0.6$ to now is supported by an independent observation at the same redshift using gas kinematics to define a baryonic Tully-Fisher relation (bTFR). In \citet{2010A&A...510A..68P}, we established for the first time the bTFR at the same redshift and found that this relation has remained invariant over the past 6\,Gyr. This indicates that the reservoir of gas that has been converted into stars during the past 6\,Gyr was already gravitationally bound to disks at z=0.6, and there is no need for external gas accretion. This result applies to rotating discs at intermediate redshift for which a closed-box evolution seems to be a reasonable scenario. In contrast, a closed box model may appear to be a crude approximation for galaxies with strong evidence of interaction such as minor \citep{2007A&A...476L..21P} or major mergers \citep{2009A&A...493..899P, 2009A&A...501..437Y, 2009A&A...496..381H, 2009A&A...496...51P, 2010arXiv1001.2564F}. In fact, around 46\% of intermediate mass galaxies at $z\sim0.6$ could have experienced a recent a major merger \citep{2009A&A...507.1313H}. Nevertheless, our result only applies to the whole intermediate mass galaxy population - which accounts for most of the gas to mass ratio of field galaxies  \citep{2006A&A...447..113L}. In this way, the transfer of gas between galaxies within this population by merger processes will not affect the global evolution of metals and gas fractions. 

The evolution at higher redshifts is more complicated to interpret because we have no way to directly trace the $z\sim2.2$ objects to their local descendants.Figure \ref{Yield_evolution} shows that galaxies go through a regime dominated by infall and outflows at $z\sim2$ to a quasi-isolated state to $z\sim0.6$. This transition phase is likely situated at $z\sim1$ around the peak of the SFR \citep{2006ApJ...651..142H}. A deeper understanding of the transition phase requires spectroscopic observations in the redshift desert z=0.8-1.8 in order to constrain the evolution of the metallicities and gas fraction and directly detect outflow and infall.\\

In the framework of the IMAGES survey, we have derived the properties of the ISM from spectroscopic data of a representative sample of 65  intermediate-mass galaxies at z=0.7. The galaxy spectra have been acquired with VLT/FORS2 at moderate spatial resolution (R=1000) and have S/N comparable to those of local SDSS. Combining this data with broad-band photometry, deep imagery from ACS/HST and integrated field spectroscopy data, we have investigated the interplay between the main ingredients of the chemical evolution: metals, gas mass, stellar mass, and the SFR.  Our conclusions are as follows:
\begin{enumerate}
 \item We have searched for the presence of large-scale outflows in the spectra by studying the kinematics and morphology of emission and absorption lines. Only five galaxies in the sample have a clear signature of strong winds with $v_{wind}\sim200-400km/s$. The fraction of galaxies with outflows represents 8\% of the population of star-forming galaxies and 4\% of the population of intermediate mass galaxies at $z\sim0.6$. Therefore, we have concluded that large-scale outflows do not play an important role in intermediate mass galaxies at this redshift. 
 \item  We have estimated the gas fraction by inverting the Schmitt-Kennicutt law and found that the gas fraction has reduced by half from z=0.6 to $z=0$. The mean gas fraction in our sample is $32\%\pm3$, compared to 15\% in the local Universe for the same stellar mass range. We have investigated the gas fraction evolution out to $z\sim2.2$ recomputing the gas fraction using the same methodology. We have estimated the evolution of the gas fraction over the past 11 Gyr in intermediate mass galaxies. The mean gas fraction has increased linearly in the past 11 Gyr at the rate of 4\% per Gyr for intermediate mass galaxies. 
 \item We have scompared the relation between metallicities and gas fractions at $z\sim0.6$ with the prediction given by a closed-box model. We find that the population of intermediate mass galaxies is in a closed-box state. We have inferred upper limits to the contribution of outflows and infalls at $z\sim0.6$ using the models of \citet{2008ApJ...674..151E}: The data is only compatible  with models with small infalls of gas $f_{i}\sim0.23\times SFR$ and negligible contributions from outflows, in agreement with the small fraction of detected outflows.
 \item We have constrained the co-evolution of the gas fraction and gas metallicity in the frame of a simple closed box chemical model. Previous analyses (eg, R08, ) did not consider the empirical evolution of Z and $f_{gas}$ explicitly. This new and more robust analysis shows that the evolution of the whole population of intermediate mass galaxies is consistent with a closed-box from z = 0.7 to now and that the exchange between intermediate mass galaxies and the IGM has been negligible during over past 6 Gyr.
\end{enumerate}

\bsp

\begin{table*}
 \centering
 \begin{minipage}{140mm}
  \caption{Basic data from galaxies in the sample: redshift and stellar mass estimated from SED fitting. Three SFRs are given: $SFR_{UV}$, $SFR_{IR}$, $SFR_{H\beta}$. The $SFR_{H\beta}$ is corrected for extinction and aperture. Galaxies with limits on extinction have lower limits on $SFR_{H\beta}$.To determine $SFR_{UV}$, we used the rest-frame 2800\AA\, luminosity from SED fitting. }
  \begin{tabular}{@{}c c c c c c c c @{}}
  \hline
 Name  & z&$\log{M*/M\odot}$& $SFR_{UV}$ &  $SFR_{IR}$ &$SFR_{H\beta}$& $SFR_{Total}$& $R_{gas}$ \\
&&[dex]&$M\odot /yr$&$M\odot /yr$&$M\odot /yr$&$M\odot /yr$&[Kpc]\\
 \hline
03.0445& 0.527&  10.74 & - &  25.67&  6.05&  25.67&  6.65$\pm$ 0.35\\
03.0523& 0.650&  10.36 & - &  99.66&  79.73&  99.66&  3.42$\pm$ 0.33\\
03.057& 0.648&  9.920 & - &  51.6&  11.62&  51.6&  2.14$\pm$ 0.25\\
03.0645& 0.527&  10.28 & - & - &  14.85&  14.85&  4.54$\pm$ 0.31\\
03.1309& 0.617&  10.92 & - &  153.7&  56.08&  153.7&  4.67$\pm$ 0.64\\
03.1349& 0.617&  10.94 & - &  97.81&  40.35&  97.81&  4.59$\pm$ 0.57\\
03.1541& 0.689&  10.39 & - &  123.878& - &  123.878&  8.51$\pm$  1.21\\
22.0429& 0.626&  10.48 & - & - &  50.2&  50.2&  3.69$\pm$ 0.19\\
22.0576& 0.890&  10.32 & - & - &  36.44&  36.44&  3.16$\pm$ 0.25\\
22.0599& 0.889&  10.62 & - & - &  161.7&  161.7&  3.47$\pm$ 0.11\\
22.0779& 0.925&  10.70 & - & - &  31.66&  31.66&  3.91$\pm$ 0.23\\
22.1064& 0.536&  10.40 & - & - &  24.61&  24.61&  2.22$\pm$    0.07\\
J033211.70-274507.6& 0.676&  9.76&  2.2 & - &  16.58&  16.58&  5.92$\pm$ 0.22\\
J033212.30-274513.1& 0.645&  10.6&  2.27&  13.61$\pm$  1.02&  18.91$\pm$  10.22&  15.88$\pm$  1.02&  7.01$\pm$ 0.20\\
J033212.39-274353.6& 0.422&  10.58&  1.68&  14.94$\pm$  1.12&  3.49$\pm$  4.89&  16.62$\pm$  1.12&  11.91$\pm$ 0.15\\
J033212.51-274454.8& 0.732&  9.67&  3.35& - &  64.07$\pm$  34.6&  64.07$\pm$  34.6&  2.98$\pm$ 0.20\\
J033213.76-274616.6& 0.678&  9.5&  1.6 & - &  16.65&  16.65&  7.16$\pm$ 0.22\\
J033214.48-274320.1& 0.546&  9.56&  1.09& - &  4.19$\pm$  6.0 &  4.19$\pm$  6.0 &  2.90$\pm$ 0.19\\
J033215.36-274506.9& 0.860&  10.46&  4.24&  5.65$\pm$ 0.54 &  59.06$\pm$  22.34&  9.89$\pm$ 0.54 &  5.56$\pm$ 0.22\\
J033217.36-274307.3& 0.647&  10.54&  4.86& - &  14.59&  14.59&  11.53$\pm$ 0.20\\
J033217.75-274547.7& 0.734&  10.3&  3.46& - &  20.79&  20.79&  9.44$\pm$ 0.24\\
J033219.32-274514.& 0.725&  10.33&  2.81& - &  47.95$\pm$  37.68&  47.95$\pm$  37.68&  7.10$\pm$ 0.22\\
J033219.96-274449.8& 0.784&  10.19&  1.97&  70.67$\pm$  1.15&  25.25&  72.64$\pm$  1.15&  10.18$\pm$ 0.28\\
J033223.06-274226.3& 0.734&  10.4&  2.92&  3.01$\pm$ 0.52 &  20.77&  5.93$\pm$ 0.52 &  15.61$\pm$ 0.22\\
J033223.40-274316.6& 0.615&  11.11&  3.07&  32.57$\pm$ 0.68 &  10.79$\pm$  7.78&  35.64$\pm$ 0.68 &  8.24$\pm$ 0.20\\
J033224.60-274428.1& 0.538&  10.05&  1.72& - &  15.44$\pm$  14.25&  15.44$\pm$  14.25&  7.37$\pm$ 0.19\\
J033225.26-274524.& 0.666&  10.5&  5.05& - &  15.84&  15.84&  5.54$\pm$ 0.20\\
J033225.46-275154.6& 0.672&  10.93&  4.82&  24.46$\pm$ 0.69 &  31.96$\pm$  13.75&  29.28$\pm$ 0.69 &  9.51$\pm$ 0.22\\
J033225.77-274459.3& 0.832&  10.14&  3.12&  7.47$\pm$ 0.58 &  24.93$\pm$  12.07&  10.59$\pm$ 0.58 &  2.77$\pm$ 0.22\\
J033226.21-274426.3& 0.495&  9.24& 0.52 & - &  9.36$\pm$  28.38&  9.36$\pm$  28.38&  3.30$\pm$ 0.17\\
J033227.36-275015.9& 0.768&  10.62&  3.49&  12.52$\pm$  1.44&  17.49$\pm$  8.15&  16.01$\pm$  1.44&  8.89$\pm$ 0.20\\
J033227.93-274353.6& 0.458&  8.94& 0.37 & - &  1.62$\pm$ 0.67 &  1.62$\pm$ 0.67 &  3.61$\pm$ 0.17\\
J033227.93-275235.6& 0.383&  10.52&  1.74&  9.90$\pm$ 0.25 &  5.05$\pm$  1.58&  11.64$\pm$ 0.25 &  10.73$\pm$ 0.15\\
J033229.32-275155.4& 0.510&  9.52&  2.09&  3.81$\pm$ 0.27 &  8.68$\pm$  6.26&  5.90$\pm$ 0.27 &  6.84$\pm$ 0.19\\
J033229.64-274242.6& 0.667&  10.95&  5.15&  122.92$\pm$  1.31&  100.72$\pm$  48.76&  128.07$\pm$  1.31&  3.19$\pm$ 0.20\\
J033229.71-274507.2& 0.737&  10.19&  2.83& - &  27.74$\pm$  21.8&  27.74$\pm$  21.8&  5.05$\pm$ 0.20\\
J033230.07-274534.2& 0.647&  10.42&  2.88& - &  15.27$\pm$  6.04&  15.27$\pm$  6.04&  6.87$\pm$ 0.20\\
J033230.57-274518.2& 0.679&  11.04&  11.24&  43.69&  33.53$\pm$  13.74&  54.93&  16.81$\pm$ 0.20\\
J033231.58-274612.7& 0.654&  9.97&  3.22& - &  43.41$\pm$  20.22&  43.41$\pm$  20.22&  11.93$\pm$ 0.22\\
J033232.13-275105.5& 0.682&  10.58&  1.03&  17.7$\pm$ 0.71 &  4.64$\pm$  8.93&  18.73$\pm$ 0.71 &  8.75$\pm$ 0.22\\
J033232.32-274343.6& 0.534&  9.56& 0.72 & - &  8.48&  8.48&  7.22$\pm$ 0.19\\
J033232.58-275053.9& 0.669&  10.46&  2.37&  4.05$\pm$ 0.31 &  16. &  6.42$\pm$ 0.31 &  8.74$\pm$ 0.22\\
J033233.00-275030.2& 0.669&  11.19&  1.11&  50.93$\pm$ 0.87 &  70.94$\pm$  34.34&  52.04$\pm$ 0.87 &  11.79$\pm$ 0.22\\
J033233.82-274410.& 0.666&  10.96& 0.83 &  3.26$\pm$ 0.40 &  15.8&  4.09$\pm$ 0.40 &  5.43$\pm$ 0.20\\
J033233.90-274237.9& 0.619&  10.58 & - &  14.31$\pm$  7.09&  10.47$\pm$  4.32&  14.31&  5.58$\pm$ 0.19\\
J033234.04-275009.7& 0.702&  10.1&  3.86& - &  9.93$\pm$  6.34&  9.93$\pm$  6.34&  8.09$\pm$ 0.20\\
J033234.88-274440.6& 0.677&  10.16& 0.26 &  21.74$\pm$ 0.67 &  28.0$\pm$  11.55&  22.0$\pm$ 0.67 &  3.74$\pm$ 0.20\\
J033234.91-274501.9& 0.665&  10.0&  2.23& - &  15.76&  15.76&  5.20$\pm$ 0.24\\
J033236.37-274543.3& 0.435&  10.34& 0.38 &  11.83$\pm$ 0.53 &  12.69$\pm$  5.23&  12.21$\pm$ 0.53 &  9.08$\pm$ 0.30\\
J033236.52-275006.4& 0.689&  10.55&  3.76&  8.71$\pm$ 0.47 &  21.88$\pm$  12.66&  12.47$\pm$ 0.47 &  11.89$\pm$ 0.20\\
J033236.74-275206.9& 0.784&  10.38&  2.61&  3.13$\pm$ 0.39 &  22.57$\pm$  10.51&  5.74$\pm$ 0.39 &  5.62$\pm$ 0.20\\
J033237.26-274610.3& 0.736&  10.29&  2.69&  2.52$\pm$ 0.33 &  53.23$\pm$  36.15&  5.21$\pm$ 0.33 &  6.80$\pm$ 0.22\\
J033237.49-275216.1& 0.423&  10.22&  1.99& - &  7.16$\pm$  2.71&  7.16$\pm$  2.71&  7.86$\pm$ 0.17\\
J033237.96-274652.& 0.619&  10.04&  3.37& - &  17.67$\pm$  13.89&  17.67$\pm$  13.89&  8.45$\pm$ 0.19\\
J033238.77-274732.1& 0.458&  10.39&  3.94&  51.17$\pm$ 0.48 &  78.58$\pm$  27.08&  55.11$\pm$ 0.48 &  2.16$\pm$ 0.17\\

\end{tabular}
\end{minipage}
\end{table*}

\begin{table*}
 \centering
 \begin{minipage}{140mm}
  \caption{ Continue}
  \begin{tabular}{@{}c c c c c c c c @{}}
  \hline
 Name  & z&$\log{M*/M\odot}$& $SFR_{UV}$ &  $SFR_{IR}$ &$SFR_{H\beta}$& $SFR_{Total}$& $R_{gas}$ \\
&&[dex]&$M\odot /yr$&$M\odot /yr$&$M\odot /yr$&$M\odot /yr$&[Kpc]\\
 \hline

J033240.32-274722.8& 0.619&  9.69&  1.14& - &  13.81$\pm$  18.14&  13.81$\pm$  18.14&  3.70$\pm$ 0.19\\
J033244.44-274819.& 0.416&  10.72&  1.4 &  9.63$\pm$ 0.24 &  10.41$\pm$  4.48&  11.03$\pm$ 0.24 &  4.80$\pm$ 0.15\\
J033245.11-274724.& 0.435&  10.76&  1.92&  20.8&  7.16$\pm$  3.73&  22.72&  3.09$\pm$ 0.17\\
J033245.63-275133.& 0.857&  10.14&  4.46& - &  13.53$\pm$  6.8 &  13.53$\pm$  6.8 &  4.23$\pm$ 0.22\\
J033245.78-274812.9& 0.534&  10.55& 0.46 &  1.59$\pm$ 0.26 &  8.54&  2.05$\pm$ 0.26 &  2.62$\pm$ 0.19\\
J033248.84-274531.5& 0.278&  9.74& 0.17 &  1.29$\pm$    0.080 &  1.06$\pm$ 0.81 &  1.46$\pm$    0.08 &  2.88$\pm$ 0.13\\
J033249.58-275203.1& 0.415&  10.53& 0.71 &  7.45$\pm$ 0.19 &  9.93$\pm$  6.34&  8.16$\pm$ 0.19 &  6.17$\pm$ 0.17\\
J033252.85-275207.9& 0.684&  10.13& 0.60 &  19.3$\pm$ 0.73 &  21.56$\pm$  10.04&  19.9$\pm$ 0.73 &  2.37$\pm$ 0.20\\
UDSF03& 0.553&  10.40 & - & - &  11.14&  11.14&  7.39$\pm$ 0.37\\
UDSF16& 0.455&  10.24 & - &  8.16& - &  8.16&  5.87$\pm$ 0.13\\
\end{tabular}
\end{minipage}
\end{table*}

\begin{table*}
 \centering
 \begin{minipage}{140mm}
  \caption{Gas fraction,  oxygen abundance and effective yield. The gas fractions have been estimated using the K-S law.  Uncertainties are from the uncertainties of the $SFR_{Total}$ and gas radius. Galaxies with limits of $SFR_{Total}$ have upper limits of gas faction. The uncertainties in metallicity are from uncertainties of extinction and emission line flux measurement.  }
  \begin{tabular}{@{}c c c c c c c@{}}
  \hline
Name  &$\log{M_*/ M\odot} $&$\log{M_{gas}/M\odot}$ & $f_{gas}$& $\epsilon _{inf} f_{gas}$& $\epsilon_{sup} f_{gas}$&$12+\log{O/H}$ \\
&[dex]&[dex]&[\%]&[\%]&[\%]&[dex] \\

 \hline

03.0445&10.74&10.19$\pm$ 0.23&22.07&15.86&16.77& 8.79$\pm$ 0.07\\
03.0523&10.36&10.45$\pm$ 0.24&55.04&23.30&24.62& 8.68$\pm$ 0.09\\
03.0570& 9.92&10.13$\pm$ 0.28&61.73&26.18&27.44& 8.87$\pm$ 0.27\\
03.0645&10.28& 9.93&30.77&-&-& 8.74$\pm$ 0.09\\
03.1309&10.92&10.66$\pm$ 0.29&35.44&26.03&27.25& 8.46$\pm$ 0.11\\
03.1349&10.94&10.51$\pm$ 0.27&27.32&21.60&22.65& 8.73$\pm$ 0.06\\
03.1541&10.39&12.10&69.19&-&-& 8.74$\pm$ 0.09\\
22.0429&10.48&10.25&37.27&-&-& 8.71$\pm$ 0.15\\
22.0576&10.32&10.12&38.47&-&-& 8.65$\pm$ 0.14\\
22.0599&10.62&10.60&48.95&-&-& 8.39$\pm$ 0.06\\
22.0779&10.70&10.13&21.04&-&-& 8.65$\pm$ 0.12\\
22.1064&10.40& 9.91&24.33&-&-& 8.53$\pm$ 0.07\\
J033211.70-274507.6& 9.77&10.03&64.67&-&-& 8.63$\pm$ 0.01\\
J033212.30-274513.1&10.63&10.06$\pm$ 0.20&20.90&12.78&13.66& 9.02$\pm$ 0.03\\
J033212.39-274353.6&10.58&10.20$\pm$ 0.18&29.50&15.30&16.41& 8.83$\pm$ 0.35\\
J033212.51-274454.8& 9.67&10.28$\pm$ 0.26&80.08&16.64&17.49& 8.54$\pm$ 0.01\\
J033213.76-274616.6& 9.50&10.08&78.96&-&-& 8.55$\pm$ 0.02\\
J033214.48-274320.1& 9.56& 9.42$\pm$ 0.37&42.22&38.04&39.34& 8.13$\pm$ 0.30\\
J033215.36-274506.9&10.46& 9.85$\pm$ 0.20&19.74&12.63&13.47& 8.73$\pm$ 0.19\\
J033217.36-274307.3&10.54&10.15&28.87&-&-& 8.90$\pm$ 0.09\\
J033217.75-274547.7&10.31&10.21&44.64&-&-& 8.68$\pm$ 0.03\\
J033219.32-274514.0&10.33&10.40$\pm$ 0.27&53.93&26.25&27.57& 8.23$\pm$ 0.07\\
J033219.96-274449.8&10.19&10.62$\pm$ 0.19&72.91&15.09&16.15& 8.70$\pm$ 0.04\\
J033223.06-274226.3&10.41& 9.95$\pm$ 0.19&25.72&14.17&15.19& 8.48$\pm$ 0.21\\
J033223.40-274316.6&11.11&10.35$\pm$ 0.19&14.60& 9.46&10.12& 9.02$\pm$ 0.03\\
J033224.60-274428.1&10.06&10.06$\pm$ 0.28&50.26&27.86&29.19& 8.27$\pm$ 0.13\\
J033225.26-274524.0&10.50&10.00&23.85&-&-& 8.76$\pm$ 0.01\\
J033225.46-275154.6&10.94&10.32$\pm$ 0.19&19.44&11.84&12.68& 8.91$\pm$ 0.04\\
J033225.77-274459.3&10.14& 9.70$\pm$ 0.23&26.43&17.38&18.42& 8.72$\pm$ 0.01\\
J033226.21-274426.3& 9.25& 9.71$\pm$ 0.53&74.26&49.77&50.79& 8.31$\pm$ 0.04\\
J033227.36-275015.9&10.63&10.12$\pm$ 0.19&23.60&13.74&14.70& 9.06$\pm$ 0.01\\
J033227.93-274353.6& 8.95& 9.18$\pm$ 0.23&63.23&21.35&22.59& 8.73$\pm$ 0.02\\
J033227.93-275235.6&10.53&10.07$\pm$ 0.18&25.75&14.06&15.07& 9.15$\pm$ 0.01\\
J033229.32-275155.4& 9.52& 9.74$\pm$ 0.19&62.36&17.99&19.24& 8.21$\pm$ 0.07\\
J033229.64-274242.6&10.96&10.51$\pm$ 0.22&26.24&16.39&17.42& 8.85$\pm$ 0.04\\
J033229.71-274507.2&10.19&10.15$\pm$ 0.27&47.57&27.07&28.39& 8.87$\pm$ 0.06\\
J033230.07-274534.2&10.43&10.04$\pm$ 0.22&28.92&17.75&18.84& 8.71$\pm$ 0.06\\
J033230.57-274518.2&11.05&10.66&28.90&-&-& 9.09$\pm$ 0.01\\
J033231.58-274612.7& 9.98&10.50$\pm$ 0.22&76.98&15.24&16.18& 8.57$\pm$ 0.03\\
J033232.13-275105.5&10.58&10.16$\pm$ 0.19&27.60&15.22&16.29& 8.84$\pm$ 0.59\\
J033232.32-274343.6& 9.56& 9.87&66.88&-&-& 8.79$\pm$ 0.12\\
J033232.58-275053.9&10.46& 9.83$\pm$ 0.19&18.86&11.67&12.49& 8.97$\pm$ 0.01\\
J033233.00-275030.2&11.19&10.55$\pm$ 0.19&18.66&11.32&12.13& 8.85$\pm$ 0.03\\
J033233.82-274410.0&10.97& 9.57$\pm$ 0.20& 3.85& 2.96& 3.15& 8.61$\pm$ 0.07\\
J033233.90-274237.9&10.58& 9.97$\pm$ 0.20&19.49&12.46&13.30& 8.83$\pm$ 0.02\\
J033234.04-275009.7&10.11& 9.95$\pm$ 0.24&40.92&23.20&24.48& 8.64$\pm$ 0.03\\
J033234.88-274440.6&10.16&10.00$\pm$ 0.21&40.95&20.01&21.30& 8.60$\pm$ 0.04\\
J033234.91-274501.9&10.00& 9.98&48.82&-&-& 8.72$\pm$ 0.02\\
J033236.37-274543.3&10.35&10.04$\pm$ 0.20&32.89&17.19&18.37& 8.76$\pm$ 0.01\\
J033236.52-275006.4&10.55&10.11$\pm$ 0.19&26.57&14.51&15.55& 8.95$\pm$ 0.03\\
J033236.74-275206.9&10.39& 9.69$\pm$ 0.20&16.50&10.89&11.62& 8.67$\pm$ 0.03\\
J033237.26-274610.3&10.30& 9.70$\pm$ 0.20&20.33&12.66&13.52& 8.86$\pm$ 0.06\\
J033237.49-275216.1&10.22& 9.84$\pm$ 0.21&29.20&17.22&18.32& 8.64$\pm$ 0.04\\
J033237.96-274652.0&10.04&10.14$\pm$ 0.26&55.30&25.38&26.70& 8.51$\pm$ 0.05\\
J033238.77-274732.1&10.39&10.15$\pm$ 0.22&36.40&20.32&21.55& 8.72$\pm$ 0.01\\
\end{tabular}
\end{minipage}
\end{table*}

\begin{table*}
 \centering
 \begin{minipage}{140mm}
  \caption{Continue}
  \begin{tabular}{@{}c c c c c c c@{}}
  \hline
Name  &$\log{M_*/ M\odot} $&$\log{M_{gas}/M\odot}$ & $f_{gas}$& $\epsilon _{inf} f_{gas}$& $\epsilon_{sup} f_{gas}$&$12+\log{O/H}$ \\
&[dex]&[dex]&[\%]&[\%]&[\%]&[dex] \\

 \hline
J033240.32-274722.8& 9.69& 9.85$\pm$ 0.35&59.30&34.67&35.96& 8.50$\pm$ 0.14\\
J033244.44-274819.0&10.73& 9.85$\pm$ 0.20&11.65& 7.95& 8.50& 8.91$\pm$ 0.01\\
J033245.11-274724.0&10.77& 9.96&13.54&-&-& 9.14$\pm$ 0.01\\
J033245.63-275133.0&10.15& 9.88$\pm$ 0.25&35.20&22.28&23.49& 8.69$\pm$ 0.05\\
J033245.78-274812.9&10.56& 9.18$\pm$ 0.23& 3.98& 3.39& 3.59& 9.01$\pm$ 0.01\\
J033248.84-274531.5& 9.75& 9.10$\pm$ 0.20&18.29&12.09&12.88& 8.76$\pm$ 0.04\\
J033249.58-275203.1&10.53& 9.82$\pm$ 0.19&16.25&10.40&11.13& 8.97$\pm$ 0.04\\
J033252.85-275207.9&10.14& 9.86$\pm$ 0.23&34.31&20.38&21.58& 8.26$\pm$ 0.03\\
UDSF03&10.40& 9.96&26.60&-&-& 8.90$\pm$ 0.07\\
UDSF16&10.24&12.01&26.88&-&-& 9.01$\pm$ 0.04\\

\end{tabular}
\end{minipage}
\end{table*}

\label{lastpage}

\end{document}